\documentclass[aip,reprint,amsmath,amssymb,amsfonts,floatfix,superscriptaddress,showkeys,showpacs]{revtex4-1}

\usepackage{dcolumn} 
\usepackage{multirow}
\usepackage[T1]{fontenc}
\usepackage{dsfont}
\usepackage{verbatim}
\usepackage{bm}
\usepackage{hyperref} 
\usepackage{cleveref} 
\usepackage{floatrow}  
\usepackage{natbib}
\usepackage{graphicx}
\usepackage{epstopdf}
\usepackage{ifpdf}
\usepackage{epsfig}
\usepackage{color}
\usepackage[normalem]{ulem}
\usepackage[usenames,dvipsnames]{xcolor}
\usepackage[caption=false]{subfig}
\usepackage[caption=false]{subfig}
\begin{document}

\title{Shear-induced bimodality and stability analysis of chiral spheroidal swimmers}

\author{Mohammad Reza Shabanniya}
\email{Corresponding author: shabannia@ipm.ir}
\affiliation{School of Physics, Institute for Research in Fundamental Sciences (IPM), Tehran 19538-33511, Iran}
\author{Amir Abbasi}
\affiliation{Fachbereich Physik, Freie Universit\"{a}t Berlin, Arnimallee 14, 14195 Berlin, Germany}
\affiliation{School of Physics, Institute for Research in Fundamental Sciences (IPM), Tehran 19538-33511, Iran}
\author{Arghavan Partovifard}
\affiliation{Institute of Theoretical Physics, Technische Universit\"{a}t Berlin, 10623 Berlin, Germany}
\affiliation{School of Physics, Institute for Research in Fundamental Sciences (IPM), Tehran 19538-33511, Iran}
\author{Ali Naji}
\affiliation{School of Nano Science, Institute for Research in Fundamental Sciences (IPM), Tehran 19538-33511, Iran}

\begin{abstract}
We study the shear-induced behavior of chiral (circle-swimming) and nonchiral swimmers in a planar channel subjected to Poiseuille (pressure-driven) flow. The swimmers are modeled as active Brownian spheroids, self-propelling with a fixed-magnitude velocity, pointing along their axis of symmetry. We consider both cases of prolate and oblate swimmers and focus primarily on swimmers that possess an intrinsic, or active, counterclockwise angular velocity, in addition to the shear-induced angular velocity they acquire within the channel flow (Jeffery orbits). The probabilistic results are established using a Smoluchowski equation within the position orientation phase space that is solved numerically. We show that the interplay between shear- and chirality-induced angular velocities, combined with the effects due to particle self-propulsion and steric particle-wall interactions, lead to an interesting range of effects, including pinning, off-centered and near-wall accumulation of swimmers within the channel. This is further accompanied by the linear stability analysis of swimmers, used to explain the probabilistic results. We discuss the qualitative differences of probabilistic results that emerge due to the different regimes of rotational dynamics of swimmers in channel flow, particularly the  shear-trapping/-escaping of prolate/oblate swimmers and the difference in latitudes of fixed points associated with each swimmer shape. Our results point to possible mechanisms for flow-driven separation of active spheroidal particles based on their chirality strength. 
\end{abstract}


\maketitle

\section{Introduction}

Understanding, controlling and utilization of active particles have attracted wide-ranging interest in applied and theoretical soft matter  \cite{ramaswamyreview,gompper_review,bechinger_review,ZottlStark_review,Romanczuk:EPJ2012,Marchetti2016_ABPMinimalModel,Saintillan2013_ComptesRendus, Saintillan2015,Aranson2013_Active,KlumppReview2019}. Among the different classes of active systems, active fluids and swimmers have been explored extensively. The behavior of swimmers in confined geometry and fluid channels have been of particular interest \cite{Baker_2019,Nash2010,KAYA2012BJ,Morales_edgeSwimming,Chilukuri,Hill_Chirality_Upstream,Kaya2009PRL,MarcosPNAS,KaturieaaoSciAdv,ZottlPoiseuille2013,ZottlPRL2012,Jiang2019,Garg2018,Ezhilan, Nili, Anand2019,Jiang2019,Alonso-Matilla2019,Bretherton,sperm-rheotaxis,spermsurface1,spermsurface2,spermsurface3,upstream2015prl,Potomkin_2017} because of their relevance to microfluidics and biomedical research  \cite{biofilmstoodley,persistent,robotic,Paxton2006_review,cargo,robotic,sperm-carrying,janusmain,Ebbens2010_Pursuit,Howse2007_Self-Motile,son_review,Rusconi2014}. In fact, most bacteria and other cells such as spermatozoa are capable of swimming through fluid media and tend to accumulate strongly near confining walls (an effect that is thought to be an important factor in the formation of biofilms \cite{son_review}) where they exhibit upstream swimming  \cite{Bretherton,sperm-rheotaxis,spermsurface1,spermsurface2,spermsurface3,upstream2015prl,Hill_Chirality_Upstream, KAYA2012BJ,Morales_edgeSwimming}. 

The effects of imposed shear flow on individual and collective motions of swimmers have also been scrutinized intensively over the last few years and various aspects of their behavior, including  hydrodynamic interactions \cite{Zottl2014}, klinotactic \cite{FaivreReview2008}, phoretic  \cite{golestanian_njp,sanojanus,Mozaffari_PF_2016,Paxton2005_Chem,ZottlStarkPhoresis} and elasto-hydrodynamic \cite{Pak2011,Elgeti_Rods_Surface} properties have been explored.  Even though most biological swimmers utilize flagella or cilia  \cite{berg1973bacteria,bergmotor,Cosson2015,upstream-goldstein} to bypass the low-Reynolds-number constraint set by the celebrated scallop theorem \cite{purcellmain}, synthetic self-propelled particles of various geometries have also been fabricated and widely used in experiments to study the nonequilibrium behaviors of active systems \cite{janusmain,Howse2007_Self-Motile,Ebbens2010_Pursuit}. 

The combined effects of swimmer self-propulsion and shear-induced torque on the distribution of active needlelike particles have been studied using a probabilistic approach based on the Smoluchowski equation on the joint position-orientation probability distribution function of swimmers in Poiseuille  \cite{Ezhilan} and Couette \cite{Nili} flows within models known as kinetic models. These models reproduce some of the key features of the shear-driven behavior of swimmers, despite the fact the swimmer-wall hydrodynamic interactions (see Refs. \cite{Mathijssen:2016a,Mathijssen:2016c} and references therein) are thus neglected . The latter include the wall-accumulation of swimmers and the upstream swimming that occur near both walls in the case of a Poiseuille flow and only near the stationary wall in the case of a Couette flow. Here, we generalize these results to the case of spheroidal particles of finite aspect ratio, including both prolate and the seldom-studied case of oblate swimmers, and consider both nonchiral and chiral  \cite{Lowen:EPJST2016,JulicherPRL} self-propelled spheroids. 

Unlike the idealized case of needlelike swimmers, spheroidal particles display the characteristic Jeffery orbits once subjected to shear flow \cite{Jeffery}. In the absence of noise, spheroidal swimmers move in deterministic trajectories set by constants of motion depending, among other parameters, on the shear and self-propulsion strengths. In the middle of channel, rotational dynamics of particles is determined by shear flow and the resulting Jeffery orbits could lead to the so-called {\em shear-trapping} of prolate swimmers across the high-shear regions of the fluid, in latitudes across the channel where shear-induced angular velocities are larger in magnitude \cite{Rusconi2014,ZottlPRL2012,ZottlPoiseuille2013,Ezhilan}. In the context of kinetic models (see Ref. \cite{Ezhilan} and references therein), the wall accumulation of swimmers (caused by self-propulsion and regulated by the persistence time of swimming) is suppressed by rotational  diffusion and shear-induced angular velocities; In Couette flow, shear-trapping opposes the self-propulsion effect and leads to less swimmer accumulation at the channel walls \cite{Nili}, while in Poiseuille flow the shear-trapping is stronger in the proximity of the walls and leads to more complicated regimes of interplay between shear and self-propulsion effects \cite{Ezhilan}.  

Swimmer with an externally or internally generated angular velocity experience a modified version of the above rotational dynamics  within the channel flow. For example, when the microchannel is (partially) subjected to an external field, e.g. magnetic field \cite{Waisbord2016,Meng2018,MRSh1}, an external torque proportional to their magnetic moment and the strength of magnetic field is applied on the swimmers. In particular, when the swimmers are subjected to Couette flow, intensifying  the applied external field beyond a certain threshold, fixes the orientation of swimmers across the channel, instead of causing modified Jeffery orbits. The resulting {\em orientational pinning} suppresses the rotational diffusion and the swimmers accumulate near the channel walls (depending on the field orientation) and even predominantly swim upstream \cite{MRSh1}. 
This paper considers the effects of imposed flow in micro-channels on chiral (circle-swimming) swimmers with intrinsic torques due to swimmer-specific mechanisms of motion that eventually lead to the rotation of swimmers with an approximately fixed angular velocity \cite{Teeffelen:PRE2008,Palacci2013,Bechinger:PRL2013,QuinckeRotor2019, Teeffelen, Reichhardt:2013, Xue:EPL2015, Mijalkov:2015, Wykes_2016, Lowen:EPJST2016,  Liebchen_2016}. We investigate whether such swimmers in Poiseuille channel flow could also exhibit orientational pinning, accumulation in a single latitude across the channel (e.g. along the channel walls) and possibly upstream swimming.

The rotational dynamics of nonchiral, spherical and prolate, swimmers in Posieuille flow has been the subject of a number of theoretical and experimental studies  \cite{Rusconi2014,ZottlPRL2012,ZottlPoiseuille2013,Ezhilan}, with emphasis on the shear-trapping and midchannel depletion. In Poiseuille flow, the shear rate is positive in the lower half, vanishes exactly in the middle, and becomes negative in the upper half of the channel (see Eqs. \eqref{vProfile} and \eqref{shear1}). Therefore, the shear-induced angular velocities of swimmers are clockwise in the lower half and counterclockwise in the upper half of the channel. As a result, in addition to Jeffery orbits \cite{Jeffery}, i.e. complete rotations of prolate and oblate swimmers in high shear regions, also termed {\em tumbling} \cite{ZottlPoiseuille2013}, swimmers in Poiseuille flow perform incomplete periodic rotations, in opposite directions, due to the change of direction of shear-induced angular velocity by crossing the midchannel line; this type of partial, reversing rotation, is termed {\em swinging} \cite{ZottlPoiseuille2013}. To analyze the deterministic trajectories of nonchiral, as well as chiral, prolate and oblate swimmers, we take a similar approach to Refs. \cite{ZottlPRL2012,ZottlPoiseuille2013}; then, we go further by performing a linear stability analysis whereby we make a connection between the deterministic swimmer dynamics and the corresponding probability distribution functions (PDFs) within the position orientation phase-space across the channel. Making such a connection between the two enables us to categorize different regimes of behavior of swimmers, depending on the swimming velocity and shear strength, yielding one of the main outcomes of this study. 

The paper is organized as follows: In Section \ref{Sec:Model}, we propose a continuum probabilistic model to study the shear-driven behavior of chiral spheroidal swimmers and also introducing the dynamic equations of motion governing the deterministic trajectories traversed by the swimmers, in the absence of thermal noise. The latter includes laying out the stability analysis of fixed points within the dynamical system. In Section \ref{Sec:Results}, we analyze the results of our continuum simulations of the Smoluchowski equation governing the probability distribution function of nonchiral active Brownian spheroids (swimmers), followed by the detailed analysis of the behavior of chiral swimmers, with the help of corresponding trajectories of swimmers, and discuss the stability and effect of fixed points and orientational pinning on the population of swimmers in different parts of the channel, in Poiseuille flow.  Section \ref{Sec:Results} ends by proposing a separation scheme for prolate swimmers with different chirality coefficients, based on different regimes of behavior and different placement of fixed points within the latitude-orientation phase space. The paper concludes in Section \ref{Sec:Conclusion} by discussing the results and limitations of our study.

\section{Model and Methods}
\label{Sec:Model}

\subsection{Active Brownian spheroids in shear flow}
\label{Subsec:ABPs}

\begin{figure}[t]
	\centering
\includegraphics[width=10.cm]{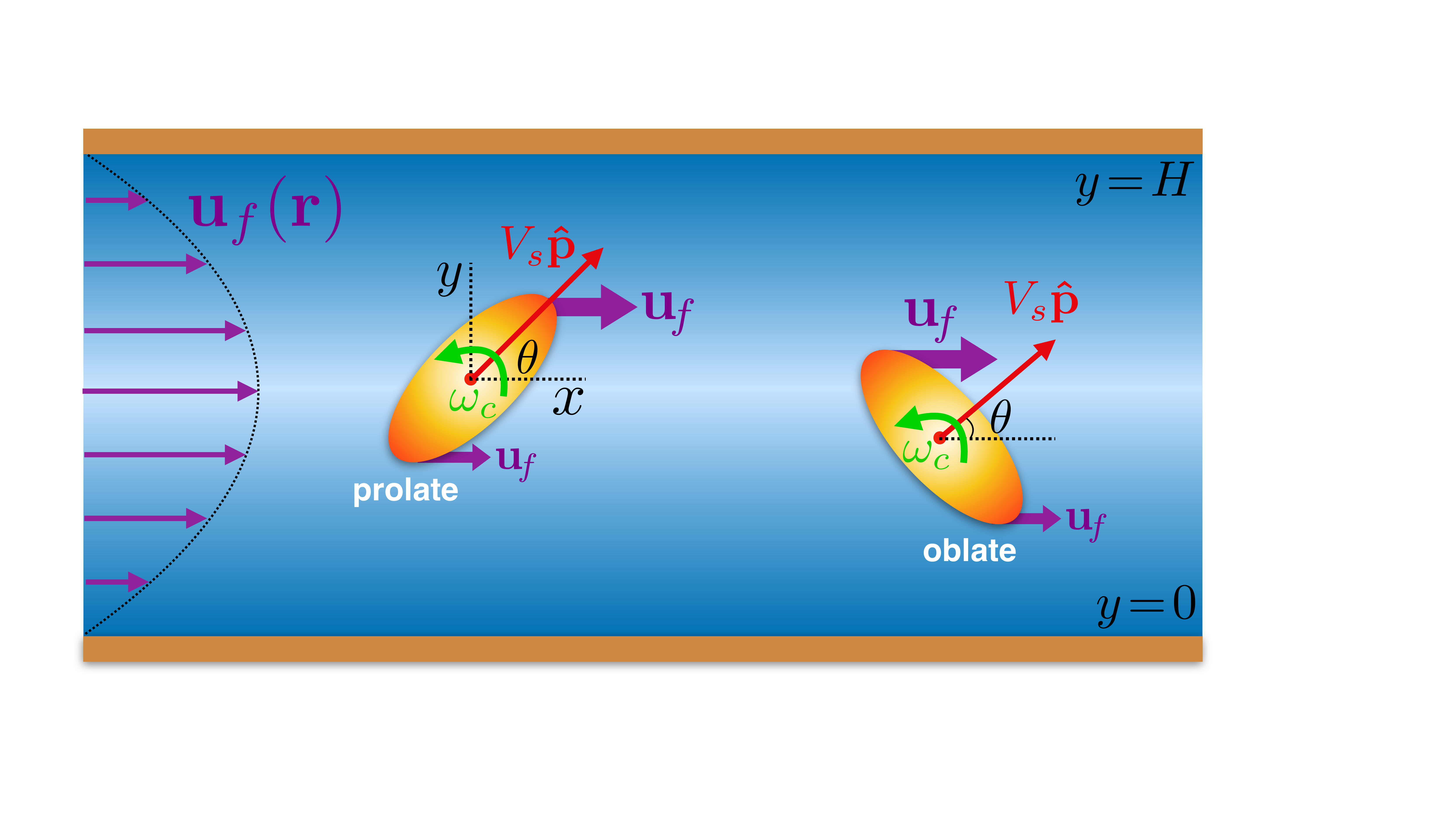}
\vspace{-1.2cm}
	\caption{Schematic view of prolate (left) and oblate (right) chiral swimmers self-propelling at constant  speed, $V_s$, along their axis of symmetry, $\hat{\mathbf{p}}$, rotating with the intrinsic angular velocity $\omega_c$ in a planar channel subjected to Poiseuille flow (shown by the parabolic profile on the left).
	}
	\label{fig:schematic}
\end{figure}

We adopt a minimal description of active Brownian particles, here referred to as swimmers, in two dimensions, as frequently discussed in the literature \cite{SaintillanPRL2008, Ezhilan, Saintillan2015, Marchetti2016_ABPMinimalModel, MRSh1}. Taking the swimmers as spheroidal particles of aspect ratio $\alpha$, we assume the swimmers to have the self-propulsion velocity, $V_s\hat{\mathbf p}$. The magnitude of swimming velocity ($V_s$) is fixed while the swim direction, being identified by the unit vector $\hat{\mathbf p}$, is taken to be aligned with the spheroidal axis of symmetry, which is the {\em major} ({\em minor}) body axis in the case of {\em prolate} ({\em oblate}) swimmers with  $\alpha>1$ ($\alpha<1$); see Fig. \ref{fig:schematic}. The spheroidal swimmers of different aspect ratio are assumed to have the same volume  ${\mathcal V}_0 = 4  \pi R_{\mathrm{eff}}^3 /3$, where $R_{\mathrm{eff}}$ is the radius of the {\em reference sphere} for $\alpha=1$. Being subject to thermal Brownian motion,  the spheroids are specified also by their bare, in-plane, rotational diffusivity, $D_R(\alpha)$, and their translational diffusion tensor (expressed in terms of parallel and perpendicular translational diffusivities, i.e., $D_\parallel(\alpha)$ and $D_\perp(\alpha)$, respectively) as
\begin{equation}
{\mathbb D}_T = D_\parallel(\alpha)\, \hat{\mathbf p}\hat{\mathbf p} + D_\perp(\alpha) \left(\mathbb{I} - \hat{\mathbf p}\hat{\mathbf p}\right).
\end{equation}

We shall consider both nonchiral and chiral swimmers. The latter, being also known as circle swimmers, will be characterized by an intrinsic torque, $\boldsymbol{{\mathcal T}}\!_c={\mathcal T}_c\, \hat{\mathbf{z}}$, pointing along the $z$-axis, identified by the unit vector $\hat{\mathbf{z}}$ normal to the plane of the flow. One can thus write the corresponding intrinsic angular velocity as 
\begin{equation}
\boldsymbol{\omega}_c= {\omega}_c\, \hat{\mathbf{z}}, 
\label{eq:w_ext}
\end{equation}
where ${\omega}_c = \mu_R(\alpha){\mathcal T}_c$ and with $\mu_R(\alpha) = D_R(\alpha)/(k_{\mathrm{B}}T)$ is the in-plane rotational mobility of the spheroids, and $k_{\mathrm{B}}T$ the ambient thermal energy scale.

The swimmer suspension is assumed to be confined in a planar channel of height $H$, with bottom and top no-slip  walls  at $\tilde y=0$ and $\tilde H$, respectively. The channel is subjected to a stationary Poiseuille flow $\mathbf{u}_f(y)=u_f(y)\hat{\mathbf{x}}$, with the `latitude' coordinate $y\in[0,H]$, and fluid velocity profile  
\begin{eqnarray}
u_f(y)  = 4y\left(1-y/H\right)  \frac{U_\text{max}}{H}
\label{vProfile}
\end{eqnarray}
hence, the shear-rate profiles
\begin{eqnarray}
\dot\gamma(y)  = 4\left(1-2 y / H\right) \frac{U_\text{max}}{H}
\label{shear1}
\end{eqnarray}
The shear-induced angular velocity, $\boldsymbol{\omega}_f$, resulting from the shear-induced torque imparted on the no-slip spheroids assumed here follows as  \cite{kim_microhydrodynamics,Dhont}
\begin{equation}
\boldsymbol{\omega}_f= \hat{\mathbf{p}} \times \big[\left(\beta(\alpha)\, {\mathbb{E}}+{\mathbb{W}}\right)\cdot{\hat{\mathbf p}}\big], 
\label{eq:w_f1}
\end{equation}
where ${\mathbb{E}}=\frac{1}{2}[\mathbf{\nabla}\mathbf{u}_f+\left(\mathbf{\nabla}\mathbf{u}_f\right)^T]$ and ${\mathbb{W}}=\frac{1}{2}[\mathbf{\nabla}\mathbf{u}_f-\left(\mathbf{\nabla}\mathbf{u}_f\right)^T]$ are the rate-of-strain and vorticity tensors, respectively, and is the $\beta(\alpha)$ Bretherton number \cite{Bretherton1962}, 
\begin{equation}
\beta(\alpha)=(\alpha^2-1)/(\alpha^2+1). 
\label{eq:beta}
\end{equation}
In the present model,  $\boldsymbol{\omega}_f$ also points along the $z$-axis, i.e., $\boldsymbol{\omega}_f={\omega}_f(y, \theta)\,\hat{\mathbf z}$, where one explicitly has \cite{ Ezhilan, Saintillan2015, MRSh1}
\begin{equation}
{\omega}_f(y, \theta)=  \frac{\dot\gamma(y) }{2} \left(\beta(\alpha) \cos2\theta-1 \right). 
\label{eq:w_f}
\end{equation}
In  Poiseuille flow, $\omega_f(y, \theta)$ changes sign through the channel centerline, $y=H/2$, from clockwise  in the bottom half of the channel to counterclockwise, $\omega_f(y, \theta)>0$,   in the top half and in all $\theta$-quadrants; however, swimmer chirality remains fixed regardless of latitude and orientation. 

When both $\omega_c$ and $\omega_f$ have the same sign, chirality enhances the angular velocity of swimmers and continually suppresses their  active features, as  the  suspension behavior tends to that of a nonactive suspension of Brownian spheroids in the limit of infinite chirality \cite{Jamali2018}. This occurs with, e.g., swimmers of clockwise chirality in the bottom half, or swimmers of counterclockwise  chirality in the top half, of Poiseuille flow. When the two angular velocities are oppositely signed and counteract one another,  the effects due to swimmer chirality  lead to significant modifications in the Jeffery orbits otherwise produced by the  shear-induced angular velocity. This can result in orientational pinning and, when combined with swimmer self-propulsion. As such, we shall restrict our analysis to the case  of counterclockwise chirality. However, an inherent mirror symmetry relative to the centerline in Poiseuille flow leads to the inclusivity of our analysis. As such, the results for swimmers of clockwise chirality in the top half can be mapped out from those obtained for swimmers of counterclockwise chirality in the bottom half of the channel, and vice versa.   

In what follows, the swimmer-wall interaction is modeled using a nearly hard, harmonic, steric interaction potential, preventing swimmers from penetrating the channel walls. The swimmers thus experience  additional (steric) translational and angular velocities, ${\mathbf u}^{(\mathrm{st})}$ and $\boldsymbol{\omega}^{(\mathrm{st})}$, as they come into contact with the channel walls; see Appendix \ref{app:y0} for further details. 

\subsection{Rescaled Smoluchowski equation}
\label{subsec:Smoluchowski_eq}

The swimmer suspension is taken to be sufficiently dilute, enabling a probabilistic description based on a non-interacting Smoluchowski equation, as frequently used in the literature \cite{Saintillan2013_ComptesRendus,Saintillan2015,gompper2006soft,doiedwards}. The steady-state Smoluchowski equation governs the joint position-orientation probability density function (PDF) of swimmers, $\Psi({\mathbf{r}},{\hat{\mathbf p}})$. We define the rotation operator as $\hat{\mathcal R}_{\hat{\mathbf p}}={\hat{\mathbf p}}\times \nabla_{\hat{\mathbf p}}$, where $\nabla_{\hat{\mathbf p}}$ represents unconstrained partial differentiation w.r.t. Cartesian components  of the orientation vector ${\hat{\mathbf p}}$ \cite{gompper2006soft,doiedwards}. Thus, the steady-state Smoluchowski equation reads \cite{MRSh1}
\begin{equation}
{\nabla}_{\mathbf{r}} \cdot \left( \mathbf{v}\Psi \right) + \hat{\mathcal R}_{\hat{\mathbf p}} \cdot \left(\boldsymbol{\omega}\Psi \right) = \nabla_{\mathbf{r}}\cdot{\mathbb D}_T\cdot \nabla_{\mathbf{r}} \Psi + D_R \hat{\mathcal R}_{\hat{\mathbf p}}^2\, \Psi.  
\label{eq:smoluchowski}
\end{equation}
The deterministic translational and angular flux velocities are respectively given as $\mathbf{v} = V_s \hat{\mathbf p} + \mathbf{u}_f(y)+{\mathbf u}^{(\mathrm{st})}$ and $\boldsymbol{\omega}\!= \! \boldsymbol{\omega}_f+\boldsymbol{\omega}_c+\boldsymbol{\omega}^{(\mathrm{st})}$. We parametrize the swim orientation, $\hat{\mathbf p}$, using the polar orientation angle, $\theta$, measured with respect to the $x$-axis (Fig. \ref{fig:schematic}), as $\hat{\mathbf p}=(\cos \theta, \sin \theta)$. Assuming spatial homogeneity along the $x$-axis, we express the joint swimmer PDF as $\Psi=\Psi(y, \theta)$. This entails that only the $y$-components of the translational flux velocities enter the Smoluchowski equation through the combination ${v}_y(y, \theta) = V_s \sin\theta +{u}_y^{(\mathrm{st})}(y, \theta)$. 
 
The numerical analysis of Eq. \eqref{eq:smoluchowski} is facilitated using a dimensionless representation obtained by rescaling the units of length and time with the lengthscale $R_{\mathrm{eff}}$ and the timescale for the rotational (Stokes) diffusivity of the reference sphere $D_{0R}=k_{\mathrm{B}}T /(8\pi\eta R_{\mathrm{eff}}^3)$, where $\eta$ is the fluid viscosity. For later use, the translational  (Stokes)  diffusivity of the reference sphere is standardly defined as $D_{0T}=k_{\mathrm{B}}T /(6\pi\eta R_{\mathrm{eff}})$. Hence, we define the rescaled quantities $\tilde y =  y/R_{\mathrm{eff}}$ and $\tilde \Psi(\tilde y, \theta)=  R_{\mathrm{eff}}\Psi(R_{\mathrm{eff}}\tilde y, \theta)$, as well as 
the rescaled translational and angular velocities as $\tilde u_y = v_y/({R_{\mathrm{eff}}D_{0R}})$ and $\tilde{  \omega} =  {{\omega}}/{D_{0R}}$. The key dimensionless parameters describing the system behavior are the rescaled channel height and the swim P\'eclet number    
\begin{equation}
\tilde H = \frac{H}{R_{\mathrm{eff}}}\,\,\ \text{and} \,\,\ Pe_s =  \frac{V_s}{R_{\mathrm{eff}}D_{0R}}, 
\end{equation}
respectively, the flow P\'eclet number
\begin{equation}
Pe_f =  \frac{4U_{\mathrm{max}}}{H D_{0R}},   
\label{shear2}
\end{equation}
and swimmer chirality coefficient  
\begin{equation}
\Gamma= \frac{{\mathcal T}_c }{k_{\mathrm{B}} T}. 
\end{equation}
In our forthcoming numerical analysis, we shall vary these  dimensionless parameters  over a wide range of values, consistent with the parameter values reported in the case of  chiral active particles in previous experimental and analytical studies \cite{Bechinger:PRL2013,Lowen:EPJST2016} (see Appendix \ref{app:parameters}).

The dimensionless Smoluchowski equation \eqref{eq:smoluchowski} can then be expressed in the $\tilde y-\theta$ domain as
\begin{eqnarray}
\label{SmolEq}
&&\frac{\partial }{\partial \tilde{y}} \big[\tilde u_y (\tilde y, \theta)\tilde{\Psi}]  + \frac{\partial}{\partial \theta} \big[ \tilde\omega(\tilde{y},\theta) \tilde{\Psi} \big]
\\
&&\qquad\,\,\,  = \frac{4}{3} \bigg( \Delta_+(\alpha)- \Delta_-(\alpha)\cos{2\theta}\bigg)\frac{\partial^2 \tilde{\Psi}}{\partial \tilde{y}^2} + \Delta_R(\alpha) \frac{\partial^2 \tilde{\Psi}}{\partial \theta^2},
\nonumber
\end{eqnarray}
where $\Delta_\pm(\alpha) =  \left(\Delta_\parallel(\alpha)\pm \Delta_\perp(\alpha)\right)/2$, and $\Delta_\parallel(\alpha)$, $\Delta_\perp(\alpha)$ and $\Delta_R(\alpha)$ are shape functions, giving the ratios between the diffusivities of the spheroids and those of the reference sphere (see Ref. \cite{MRSh1}  for relevant   expressions) 
\begin{equation}
\Delta_{\parallel,\perp}(\alpha) =  \frac{D_{\parallel,\perp}(\alpha)}{D_{0T}}\,\,\,\,{\textrm{and}}\,\,\,\,\Delta_R(\alpha) =  \frac{D_R(\alpha)}{D_{0R}}. 
\end{equation}
The rescaled  deterministic fluxes in Eq. \eqref{SmolEq} are 
\begin{equation} 
\tilde u_y (\tilde y, \theta) = {Pe}_s \sin{\theta}+\tilde {u}_y^{(\mathrm{st})}(\tilde y, \theta), 
\label{eq:v_y}
\end{equation}
and $\tilde\omega(\tilde{y},\theta) = \tilde\omega_f(\tilde{y},\theta)+\tilde\omega_c+\tilde{\omega}^{(\mathrm{st})}(\tilde{y},\theta)$, where 
\begin{align}
&\tilde\omega_f(\tilde y, \theta) = \frac{Pe_f}{2}\left( 1- 2\frac{\tilde y}{\tilde H}\right)\bigg(\beta(\alpha) \cos 2 \theta -1\bigg), 
\label{eq:Gamma_f}
\\
&\tilde\omega_c =\Delta_R(\alpha)\Gamma,
\label{eq:Gamma_ext}
\end{align}
and $\tilde {u}_y^{(\mathrm{st})}(\tilde y, \theta)$ and $\tilde{\omega}^{(\mathrm{st})}(\tilde{y},\theta)$ are given explicitly in Appendix \ref{app:y0}. However, in this work, we are not concerned with exact description of the near-wall behavior; we shall return to this point in Section \ref{subsec:MigrationPatterns}. 

The Smoluchowski equation \eqref{SmolEq}  is  solved using finite-element methods \cite{MRSh1}, within the position-orientation domain taken conventionally as $\tilde y\in [0, \tilde H]$ and $\theta\in [-\pi/2, 3\pi/2)$, while supplemented with periodic boundary conditions over $\theta$ and the normalization condition 
\begin{equation}
\int_{-\pi/2}^{3\pi/2}\int_0^{\tilde H}{\mathrm{d}}\tilde y\,{\mathrm{d}}\theta\, \tilde \Psi(\tilde y, \theta)=1.
\label{eq:norm}
\end{equation}

\subsection{Deterministic swimmer trajectories and fixed points}
\label{subsec:Deterministic}

Our probabilistic analysis of spheroidal swimmers will be supplemented by deterministic swimmer trajectories obtained in the absence of Brownian noises to gain further insight into the shear- and chirality-induced behavior of swimmers within the channel.  The deterministic trajectories are obtained by solving the set of dynamical equations $\tilde{\dot{y}}(\tilde t)= \tilde u_y \left(\tilde y(\tilde t), \theta(\tilde t)\right)$  and $\tilde{\dot{\theta}}(\tilde t) = \tilde\omega\left(\tilde y(\tilde t), \theta(\tilde t)\right)$. As we investigate in Sections  \ref{SubSec:Non} and \ref{SubSec:ChiralPoiseuille}, these trajectories either correspond to closed orbits around fixed points within the latitude-orientation, i.e., $\tilde y-\theta$ phase space or periodic rotations that cover the whole range of swimmer orientations.

The fixed points within the latitude-orientation phase space, with the coordinates $(  \tilde y_\ast, \theta_\ast)$, are determined by the conditions $\tilde{\dot{y}} = \tilde u_y( \tilde y_\ast, \theta_\ast) = 0$ and $\dot{\theta} = \tilde \omega(  \tilde y_\ast, \theta_\ast) = 0$. Thus, using Eqs. \eqref{eq:v_y} -- \eqref{eq:Gamma_ext}, we can write, 
\begin{align}
&Pe_s \sin \theta_\ast +  \tilde u_y^{\mathrm{(st)}} = 0, 
\label{eq:FPeq1}
\\
&\frac{Pe_f}{2}\left( 1- 2\frac{\tilde y_\ast}{\tilde H}\right)\bigg(\beta(\alpha) \cos 2 \theta_\ast -1\bigg) + \Delta_R(\alpha)\Gamma = 0
\cdot
\label{eq:FPeq2}
\end{align}
On the standard linearization level \cite{Nayfeh2008}, small perturbations $\upsilon =  \tilde y - \tilde y_\ast $ and $\vartheta =\theta - \theta_\ast$  around a given fixed point $( y_\ast, \theta_\ast)$ are governed by the  equations $\dot{\mathbf w}= {\mathbb J}\,{\mathbf w}$, where ${\mathbf w}^T=[\upsilon\,\,\,\, \vartheta]$  and 
\begin{equation}
{\mathbb J}  = 
\begin{bmatrix}
 \frac{\partial \tilde u_y}{\partial \tilde y} & \frac{\partial \tilde u_y}{\partial \theta} 
 \\ \\
\frac{\partial \tilde\omega}{\partial \tilde y } & \frac{\partial \tilde\omega}{\partial \theta }
\end{bmatrix}_{\tilde y_\ast,\theta_\ast}
 \label{eq:Jacobian}
\end{equation}
is the Jacobian matrix. The general form of the derivatives in Eq. \eqref{eq:Jacobian} can be written by using Eqs. \eqref{eq:v_y} -- \eqref{eq:Gamma_ext} as
\begin{equation}
\begin{aligned}
& \frac{\partial\tilde u_y}{\partial \tilde y} = \frac{\partial\tilde u^{\mathrm{(st)}}_y}{\partial \tilde y}, \\
& \frac{\partial \tilde u_y}{\partial \theta} = Pe_s \cos\theta + \frac{\partial\tilde u^{\mathrm{(st)}}_y}{\partial \theta},\\
& \frac{\partial \tilde\omega }{\partial\tilde y} =  \frac{Pe_f }{\tilde H} \left(1-\beta(\alpha) \cos 2\theta \right) \\
& \frac{\partial \tilde\omega}{\partial\theta} = 
 - Pe_f \beta(\alpha) \sin2\theta \left(1-2\frac{\tilde y}{\tilde H}\right)
,
\label{eq:JacobianDevs}
\end{aligned}
\end{equation}
Where the derivatives of steric velocities are given by,
\begin{equation}
\begin{aligned}
&  \frac{\partial\tilde u^{\mathrm{(st)}}_y}{\partial \tilde y} = -\kappa, \\
&  \frac{\partial\tilde u^{\mathrm{(st)}}_y}{\partial \theta} = \frac{\kappa \alpha^{2/3}\sin 2\theta}{\left(\sin^2 \theta + \alpha^{-2} \cos^2 \theta \right)}\left(\frac{\alpha^2-1}{\alpha^2}\right)
\cdot
\label{eq:stDevs}
\end{aligned}
\end{equation}
The eigenvalues of Jacobian matrix \eqref{eq:Jacobian}, $\lambda_y$ and $\lambda_\theta$, determine the {\em stability} of the fixed point(s), $(\tilde y_\ast,\theta_\ast)$; these are calculated by using the equation,
\begin{equation}
\left(\lambda - \frac{\partial\tilde u_y}{\partial \tilde y}\right) \left(\lambda - \frac{\partial \tilde\omega}{\partial\theta} \right) -
 \frac{\partial \tilde u_y}{\partial \theta} \frac{\partial \tilde\omega }{\partial\tilde y}  = 0
,
\label{eq:lambdas}
\end{equation}
where the derivatives are calculated at the fixed point, i.e., by putting $(\tilde y_\ast,\theta_\ast)$ in Eq. \eqref{eq:JacobianDevs}. 

We will discuss the deterministic trajectories of nonchiral swimmers and use the above deterministic formulation to list the fixed points of the system, associated with nonchiral and chiral swimmers, and discuss their stability in section \ref{susubsec:NonChiralDet} and Appendix \ref{App:ChiralStability}, respectively. The consequence of these deterministic results for the probabilistic analysis of nonchiral and chiral swimmers are discussed in Sections \ref{susubsec:NonChiralProb} and \ref{SubSec:ChiralPoiseuille}, respectively.

\section{Results}
\label{Sec:Results}

\begin{figure*}[t]
	\begin{center}
		\begin{minipage}[b]{0.3\textwidth}
			\begin{center}
				\includegraphics[height=0.7\linewidth]{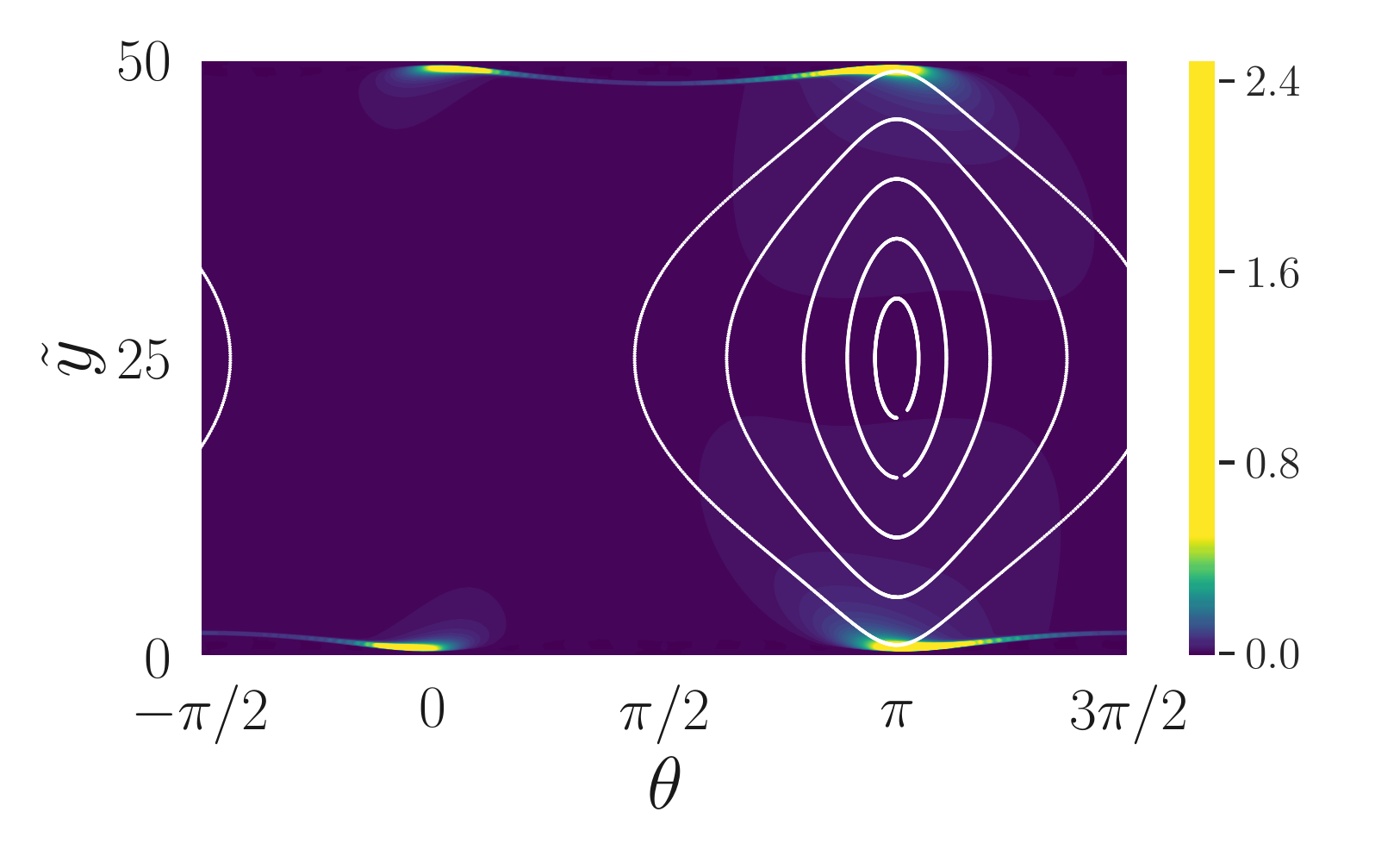}
				\vskip-1mm{(a) $\alpha=3$ and $Pe_f=5$}\label{nonchiralpoiseuillep1}
			\end{center}  
		\end{minipage}\hskip5mm
		\begin{minipage}[b]{0.3\textwidth}
			\begin{center}
				\includegraphics[height=0.7\linewidth]{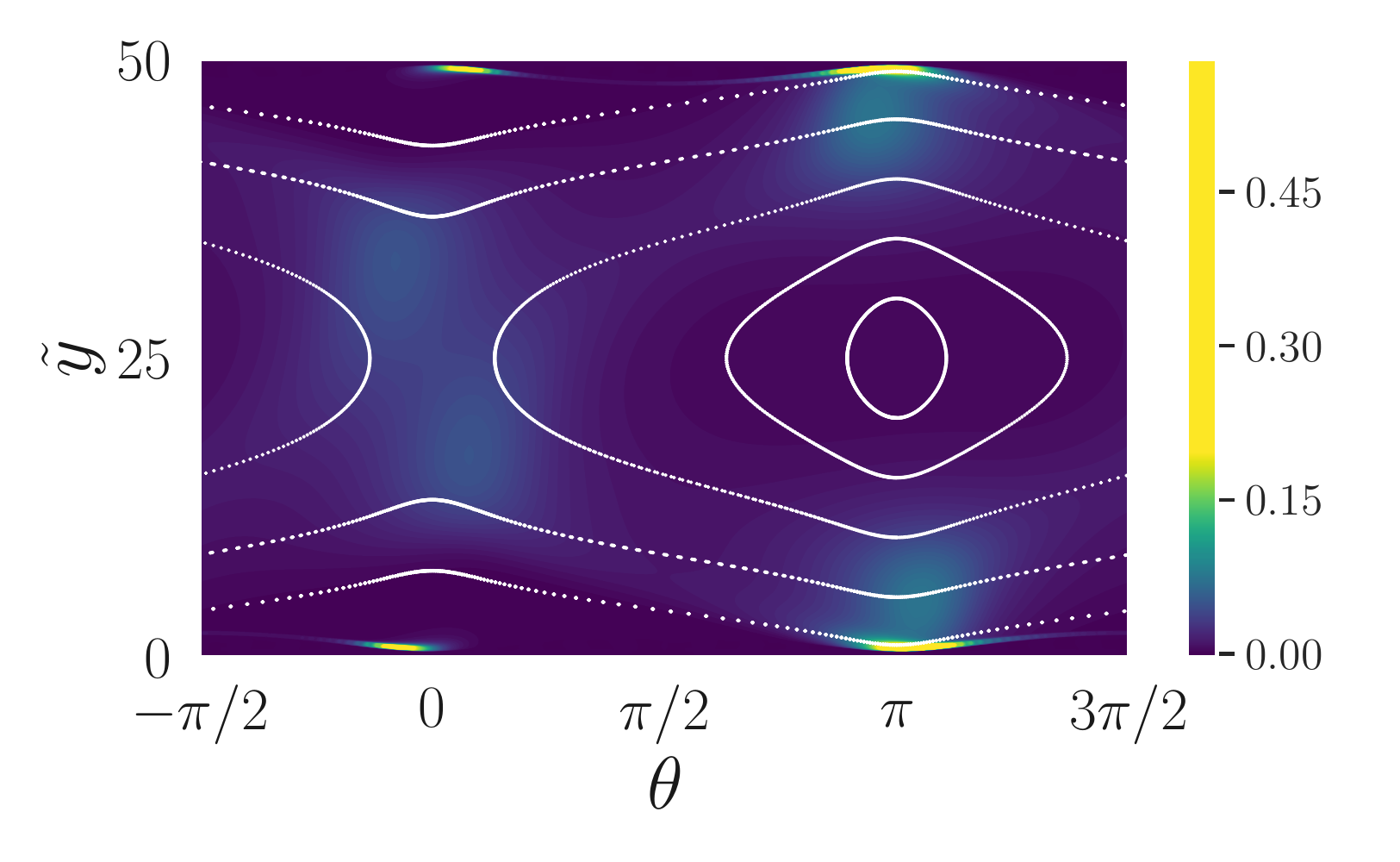}
				\vskip-1mm{(b) $\alpha=3$ and $Pe_f=20$}\label{nonchiralpoiseuillep2}
			\end{center}  
		\end{minipage}\hskip5mm
		\begin{minipage}[b]{0.3\textwidth}
			\begin{center}
				\includegraphics[height=0.7\linewidth]{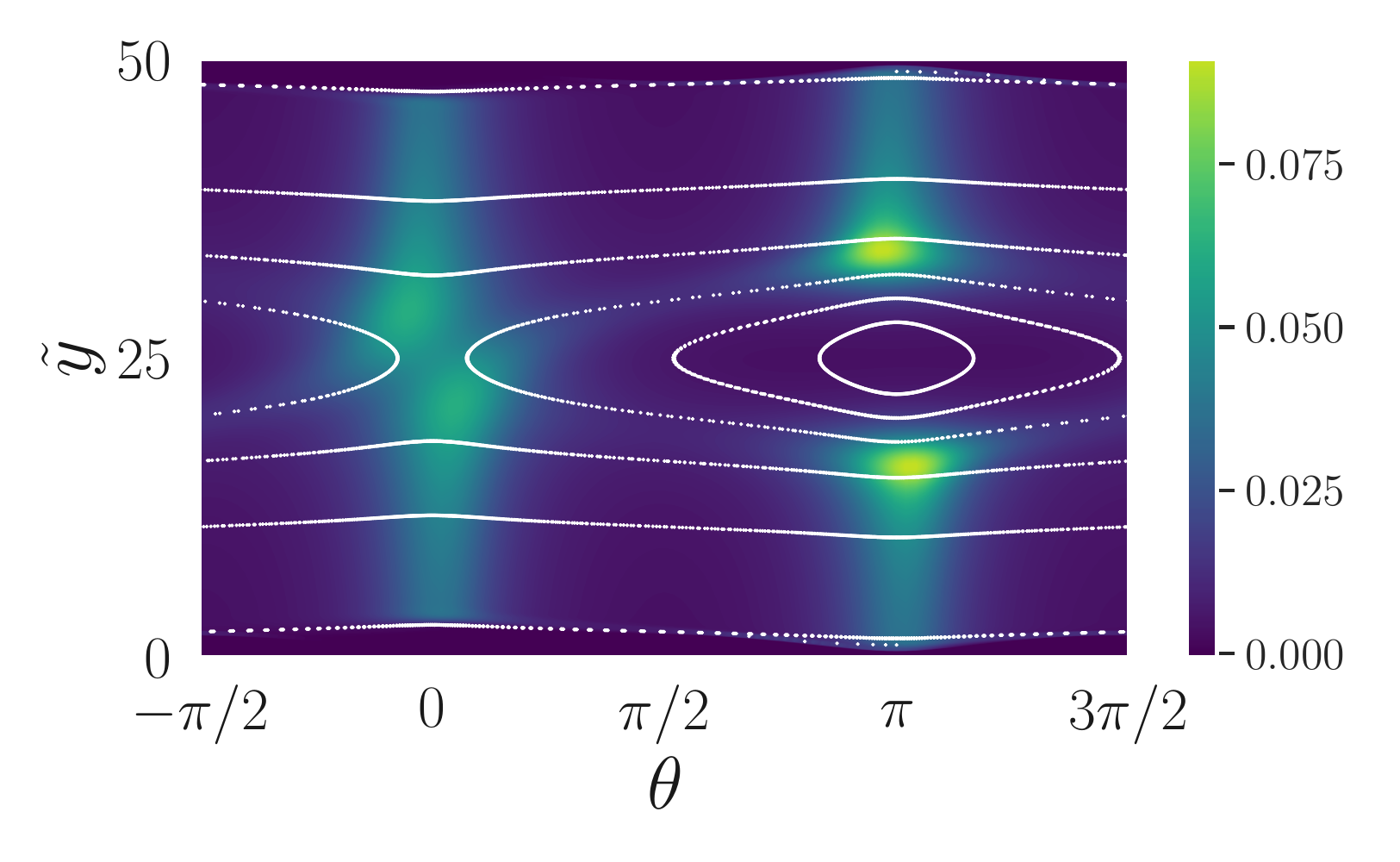}
				\vskip-1mm{(c) $\alpha=3$ and $Pe_f=100$}\label{nonchiralpoiseuillep3}
			\end{center}  
		\end{minipage}\\
		\begin{minipage}[b]{0.3\textwidth}
			\begin{center}
				\includegraphics[height=0.7\linewidth]{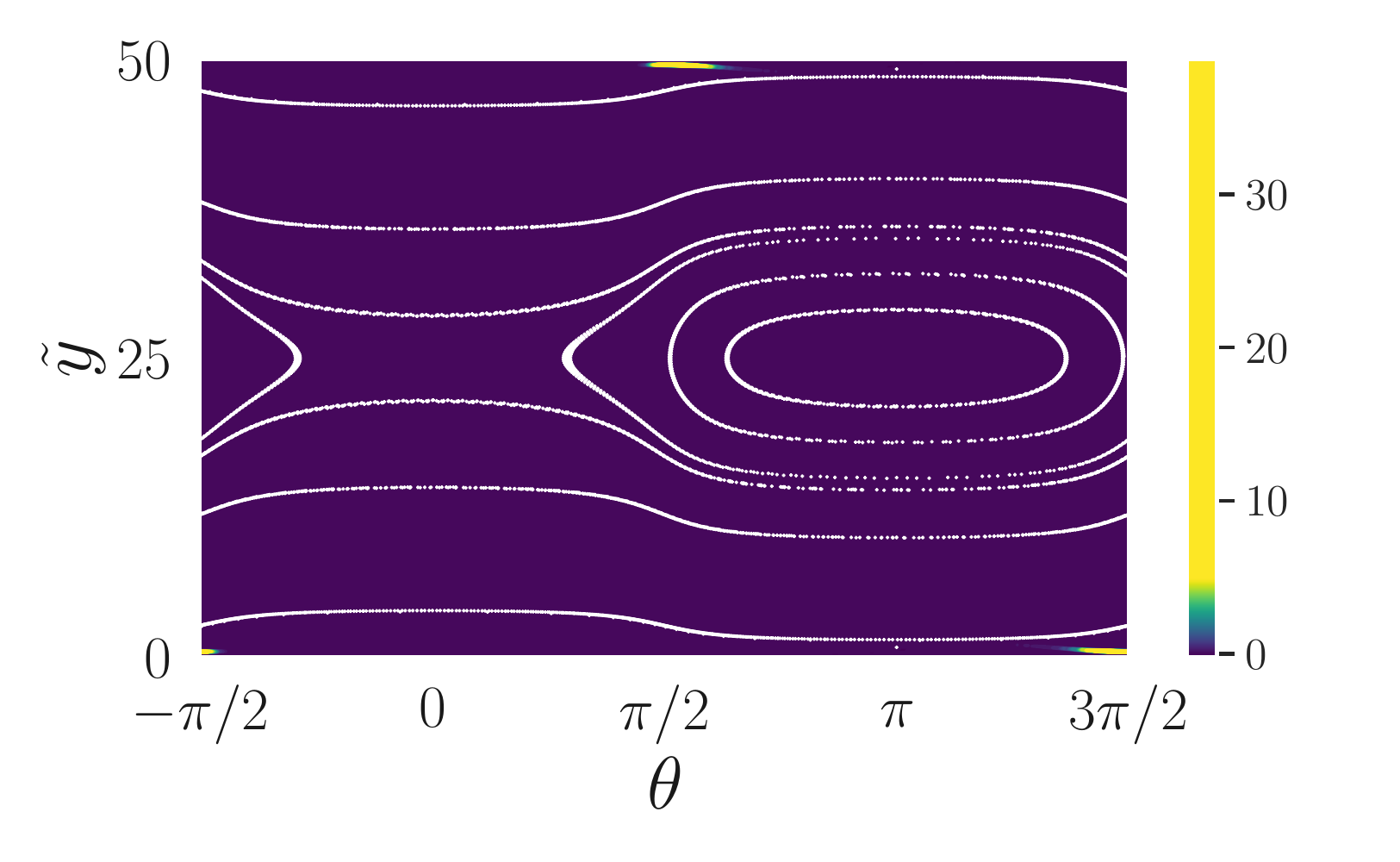}
				\vskip-1mm{(d) $\alpha=1/3$ and $Pe_f=100$}\label{nonchiralpoiseuillep4}
			\end{center}  
		\end{minipage}\hskip5mm
		\begin{minipage}[b]{0.3\textwidth}
			\begin{center}
				\includegraphics[height=0.7\linewidth]{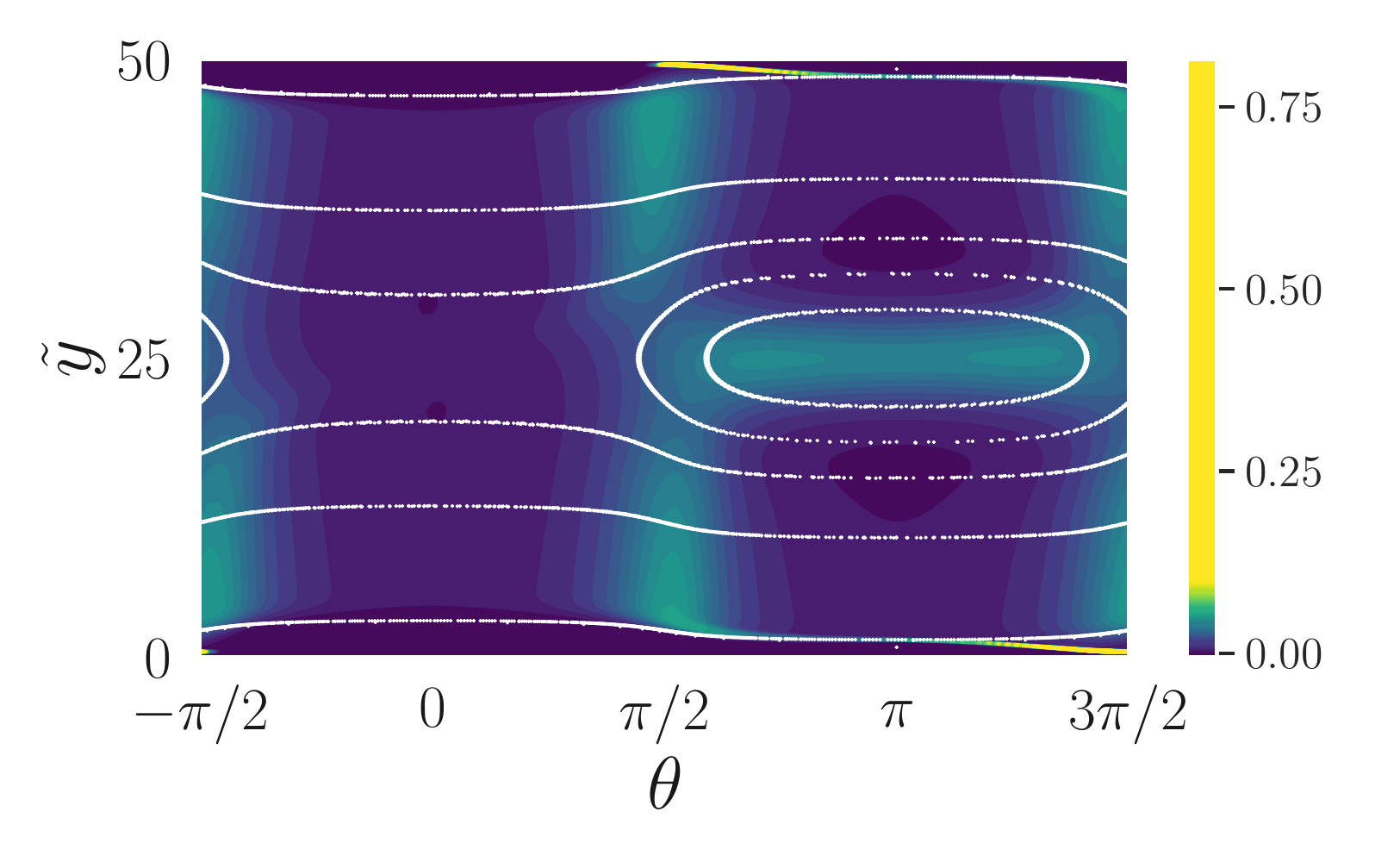}
				\vskip-1mm{(e) $\alpha=1/3$ and $Pe_f=150$}\label{nonchiralpoiseuillep5}
			\end{center}  
		\end{minipage}\hskip5mm
		\begin{minipage}[b]{0.3\textwidth}
			\begin{center}
				\includegraphics[height=0.7\linewidth]{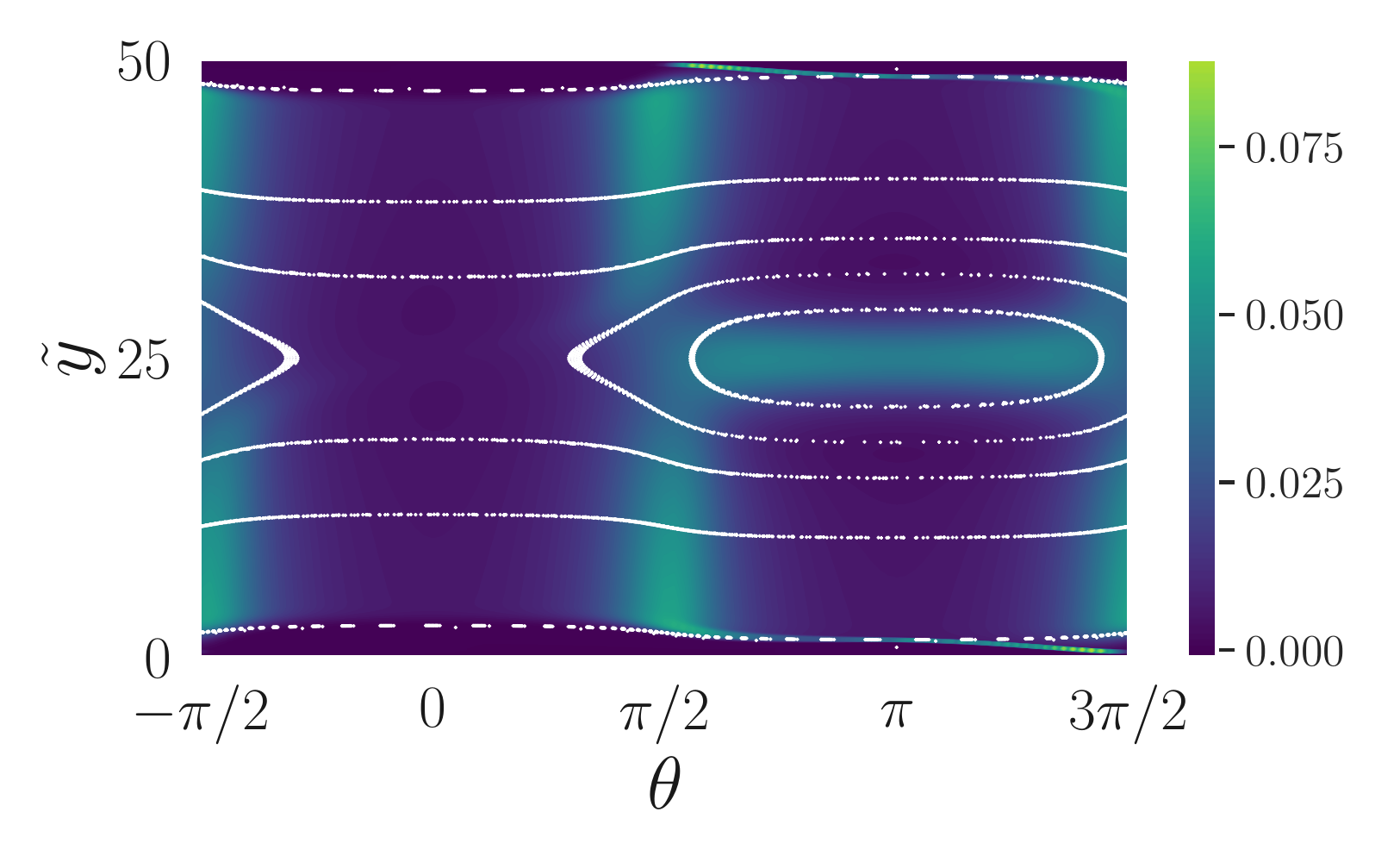}
				\vskip-1mm{(f) $\alpha=1/3$ and $Pe_f=200$}\label{nonchiralpoiseuillep6}
			\end{center}
		\end{minipage}
	\end{center}
	\vskip\baselineskip
	\vskip-4mm
	\caption{Probability distribution function, $\tilde \Psi(\tilde y, \theta)$, of prolate (a)--(c) and oblate (d)--(f) swimmers, with aspect ratios $\alpha=3$ and $\alpha=1/3$, respectively. The swimmers have the same swimming P\'eclet number $Pe_s=50$, and are subject to Poiseuille flow, with a variety of flow P\'eclet numbers, $Pe_f$, in a channel with the fixed height $\tilde{H}=50$. In all panels, the white dashed curves show the deterministic trajectories of swimmers, for the same parameters (see Section \ref{subsec:Deterministic}).} 
	\label{fig:pef_poiseuille}
\end{figure*}

\subsection{Non-chiral swimmers in Poiseuille flow}
\label{SubSec:Non}

We first consider the case of non-chiral, active Brownian spheroids (swimmers), as described in Section \ref{Subsec:ABPs}, subjected to Poieuille flow. Therefore, the oblate and prolate active particles are subjected to a shear-induced external torque, a function of both latitude ($\tilde y$) and swimmer orientation ($\theta$), giving rise only to the flow-induced angular velocity of Eq. \eqref{eq:Gamma_f}. In the foregoing discussion, we explore different aspects of the probabilistic results for non-chiral swimmers and then connect these results to the deterministic stability analysis and swimmer trajectories.

\subsubsection{Probabilistic results: shear-trapping and shear-escaping}
\label{susubsec:NonChiralProb}

In this section, we analyze the probabilistic behavior of nonchiral swimmers subject to  Poiseuille flow. In order to describe the effects of tumbling and swinging motion on the PDF, $\tilde \Psi (\tilde y , \theta)$, of spheroidal swimmers across the channel, in Fig. \ref{fig:pef_poiseuille}, we overlay the deterministic trajectories of prolate and oblate swimmers on their corresponding PDFs. The deterministic trajectories are obtained by solving the set of dynamical equations 
\begin{align}
\left\{\begin{array}{ll}
&\tilde{\dot{y}}(\tilde t)= \tilde u_y \left(\tilde y(\tilde t), \theta(\tilde t)\right)\,,  \\ \\
&\dot{\theta}(\tilde t) = \tilde\omega\left(\tilde y(\tilde t), \theta(\tilde t)\right)
\,\cdot
\end{array}\right.
\label{eq:trajectory1}
\end{align}
The dotted lines in panels a--f of Fig. \ref{fig:pef_poiseuille} depict the deterministic trajectories of spheroidal swimmers, with swimming P\'eclet number $Pe_s=50$, subject to shear flows of varying flow P\'eclet numbers. Both prolate ($\alpha=3$) and oblate ($\alpha=1/3$) swimmers have either open or closed trajectories in $\tilde y - \theta$ plane, depending on the latitude across the channel; the open trajectories correspond to tumbling motion, while the closed trajectories represent swinging motion. Here, we discuss the general trends regarding the PDF of prolate and oblate, nonchiral swimmers. The connection with deterministic analysis will become clear in Section \ref{susubsec:NonChiralDet}. 

Panel a of Fig. \ref{fig:pef_poiseuille} shows that prolate nonchiral swimmers under weak Poiseuille flow, i.e. $Pe_f =5$, exhibit wall-accumulation and their PDF is vanishingly small in the middle of the channel, for all swimmer orientations. The same observation is true for oblate swimmers, as can be seen in panel d of Fig. \ref{fig:pef_poiseuille}, even under significantly stronger fluid flow with P\'eclet number $Pe_f =100$; this stronger wall-accumulation of oblate swimmers was previously observed in the case of Couette flow \cite{MRSh1}, as well.

In the case of prolate swimmers, panels b and c of Fig. \ref{fig:pef_poiseuille} show the effect of increasing the flow strength to $Pe_f=20$ and $Pe_f = 100$, respectively. Increasing the flow P\'eclet number detaches the swimmers from the walls due to the raising shear-induced angular velocity. The detached prolate swimmers follow the deterministic trajectories, as shown in panels b and c of Fig. \ref{fig:pef_poiseuille}. Most notably, the closed trajectories around the point $(\tilde H/2, \pi)$ within the phase space keep the prolate swimmers away from this point, while pushing them toward the high-shear regions and away from the midchannel line ($\tilde y=\tilde H/2$).

Panels e and f of Fig. \ref{fig:pef_poiseuille} similarly show that by increasing the flow P\'eclet number to $Pe_f=150$ and $Pe_f = 200$, respectively, leads to the detachment of swimmers form the channel walls. However, in contrast to the case of prolate swimmers, discussed above, oblate swimmers occupy the region of phase space within the closed trajectories around the point $(\tilde H/2, \pi)$ and, in this way, escape from the high shear regions toward the midchannel line.  The significance of the point $(\tilde H/2, \pi)$ becomes clear when discussing the deterministic trajectories in Section \ref{susubsec:NonChiralDet}.

\subsubsection{Determinsitic results: Fixed points and the separatix between closed and periodic orbits}
\label{susubsec:NonChiralDet}

In order to shed light on the probabilistic dynamics of nonchiral spheroidal swimmers, as laid out in Sec. \ref{susubsec:NonChiralProb}, we first perform the stability analysis of deterministic swimmer trajectories. Equations \eqref{eq:FPeq1} and \eqref{eq:FPeq2} determine the fixed points of the logistic map within the ($\tilde y, \theta$) phase space. As such, nonchiral  swimmers ($\Gamma = 0$) have two {\em central fixed points}, i.e. located on the channel centerline, given as $(\tilde H/2 , 0)$ and $(\tilde H/2, \pi)$. Stability of these fixed points is analyzed by the procedure outlined in Section \ref{subsec:Deterministic} and eigenvalues of the Jacobian matrix, Eq. \eqref{eq:Jacobian}, at the two central fixed points are calculated as,
\begin{align}
\left\{\begin{array}{ll}
&\lambda = \pm \sqrt{{Pe_s Pe_f \left(1 - \beta\right)}/{\tilde H}} \,; \quad \theta = 0\,,  \\ \\
&\lambda = \pm i \sqrt{{Pe_s Pe_f \left(1 - \beta\right)}/{\tilde H}} \,; \quad \theta = \pi
\,\cdot
\end{array}\right.
\label{eq:MidChLamPi_nonchiral}
\end{align}
Therefore, the fixed point $(\tilde H/2 , 0)$ is a {\em saddle-point} while $(\tilde H/2 , \pi)$ is a {\em center} \cite{Nayfeh2008}. As such, there exists a swimmer phase trajectory that connects the saddle-point, $(\tilde H/2 , 0)$, back to itself, i.e., a rotation over the whole $[-\pi/2,3\pi/2)$ orientation space; this is a {\em homoclinic orbit} or the {\em separatix} \cite{Nayfeh2008}: The periodic, swinging trajectories around the center within the phase space are separated from the periodic, tumbling trajectories by the separatix (see also Refs. \cite{ZottlPRL2012,ZottlPoiseuille2013}); the latter correspond to full rotations covering the whole $[-\pi/2,3\pi/2)$ orientation space, while the former represent oscillations of swimming orientation around $\theta = \pi$ (see Section \ref{SubSec:Non}). 

Rotational and translational diffusive motion causes the swimmers to cross the above mentioned separatix within the phase space by the small changes of constants of motion \cite{ZottlPoiseuille2013} (see Appendix \ref{App:Cs}). For any given swimming P\'eclet number, $Pe_s$, the separatix shrinks by increasing the flow P\'eclet number, $Pe_f$, as can be seen by looking at swimmer trajectories in panels a--c and d--f of Fig. \ref{fig:pef_poiseuille}, for prolate and oblate swimmers, respectively. One can show that the height of the separatix is proportional to the fraction of swimming speed to flow strength \cite{ZottlPoiseuille2013} (i.e., proportional to $Pe_s/Pe_f$). The panels a--c (d--f)  of Fig. \ref{fig:pef_poiseuille} show that increasing the flow P\'ecelt number leads to the detaching of prolate (oblate)  swimmers from the channel walls due to increasingly stronger shear-induced angular velocities. Indeed, the separatix detaches from the channel walls when the flow strength surpasses a critical value, i.e., when $Pe_f > Pe_f^\ast$, as detailed in Appendix \ref{App:Cs}. As an example, for prolate ($\alpha=3$) swimmers with $Pe_s = 50$, we get $Pe_f^\ast \simeq9.3 $, and the above statement about decreased wall accumulation is seen by comparing panels a and b of Fig. \ref{fig:pef_poiseuille} which correspond to $Pe_f < Pe_f^\ast$ and $Pe_f > Pe_f^\ast$ respectively. Thus, intensifying the fluid flow and passing the threshold $Pe_f = Pe_f^\ast$, wall accumulation dramatically decreases. Therefore, we term the regime  $Pe_f > Pe_f^\ast$ as the {\em shear-dominant regime} and  $Pe_f \leq Pe_f^\ast$ as the {\em swimming-dominant regime}, where in the latter wall accumulation is the dominant effect inside the channel.

When swimmers are detached from the walls, they follow the deterministic trajectories (white dotted lines) and eventually reach the middle of the channel. However, prolate swimmers have lower PDFs inside the separatix, when compared to oblate swimmers (compare panels b--c of Fig. \ref{fig:pef_poiseuille} with the panels e--f of the same figure); since the shear rate is minimized inside the separatix, the said behaviors are indicative of {\em shear-trapping} of prolate swimmers \cite{Rusconi2014,ZottlPRL2012,ZottlPoiseuille2013,Ezhilan} and {\em shear-escaping} of oblate swimmers, respectively. These effects are probed further below. 

\begin{figure}[t!]
	\begin{center}
		\begin{minipage}[b]{\textwidth}
			\begin{center}
				\includegraphics[width=\textwidth]{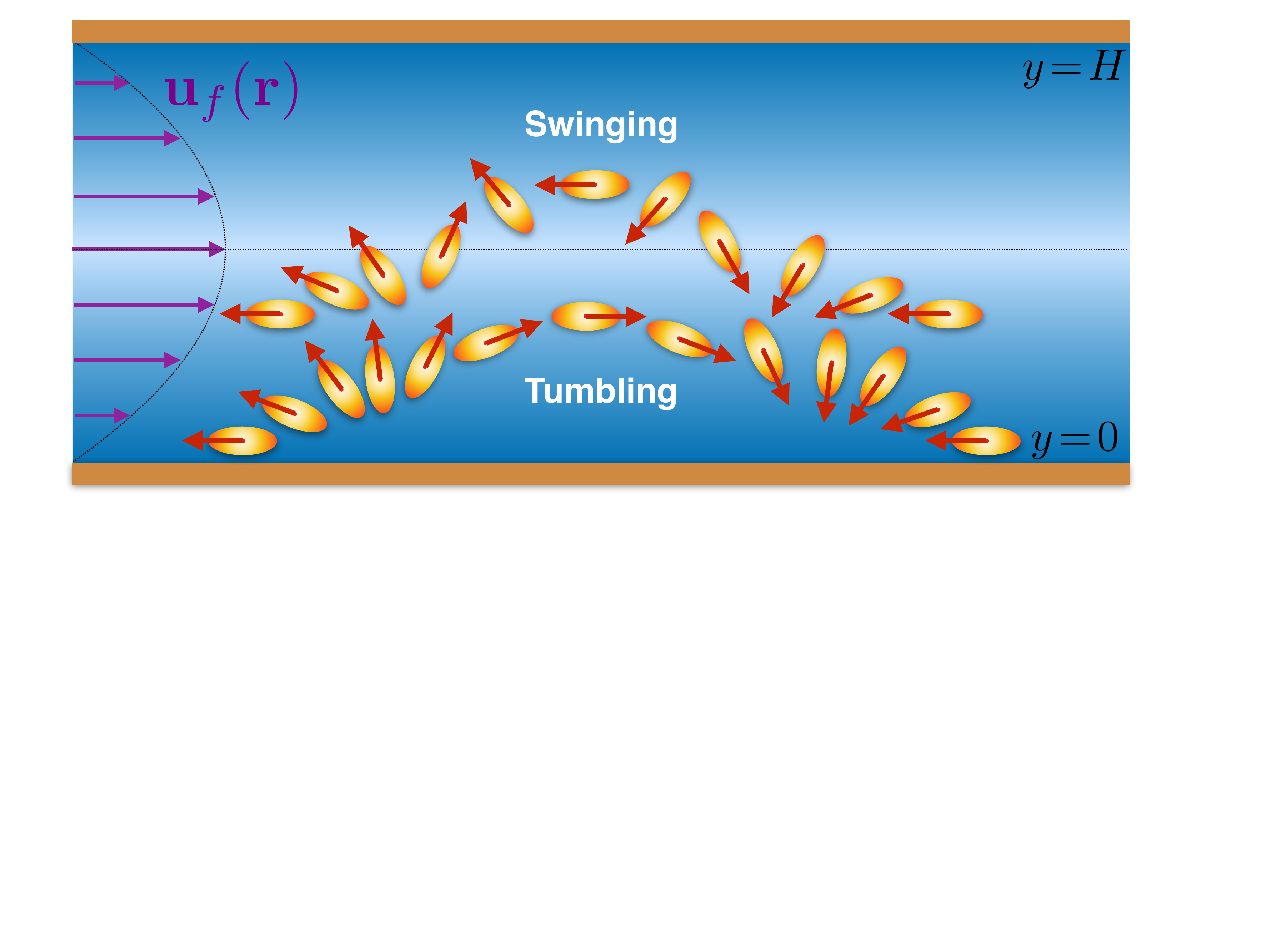}
				\vskip-30mm 
				{(a) $\alpha=3$}    
				\label{fig:prolate_poiseuille}
			\end{center}
		\end{minipage}
		\\
		\begin{minipage}[b]{\textwidth}  
			\begin{center} 
				\includegraphics[width=\textwidth]{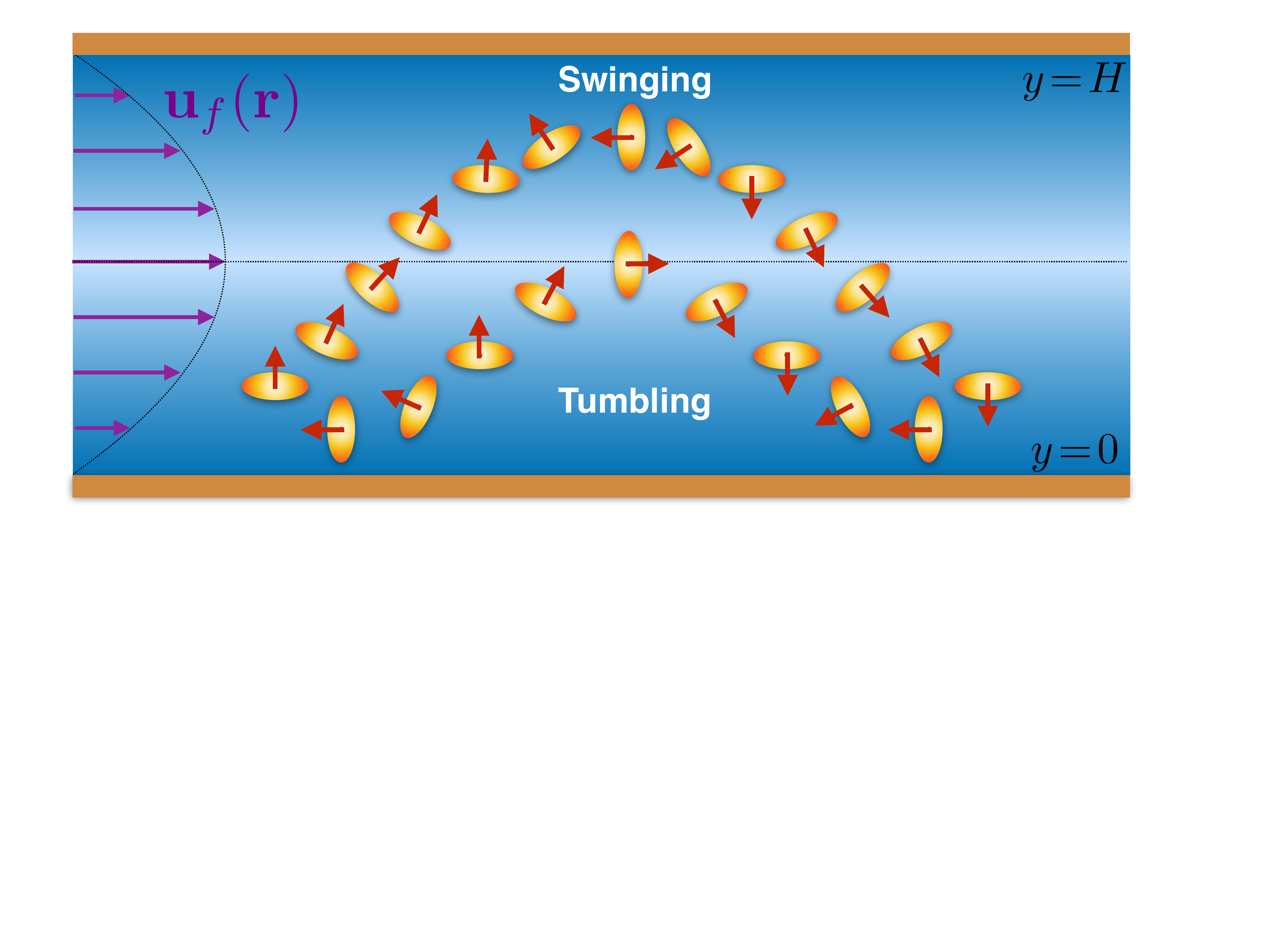}
				\vskip-30mm  
				{(b) $\alpha=1/3$}
				\label{fig:oblate_poiseuille}
			\end{center}
		\end{minipage}
		\vskip-4mm\caption{Simulated trajectories of prolate (a) and oblate (b) swimmers with aspect ratios  $\alpha=3$ and  $\alpha=1/3$, respectively, with the same swimming P\'eclet number $Pe_s=50$, subject to Poiseuille flow with $Pe_f=100$. Nonchiral prolate and oblate swimmers in Poiseuille flow perform both types of swinging motion in the middle of the channel and tumbling motion closer to the walls.} 
		\label{fig:swing_tumble}
	\end{center}
\end{figure}

For illustration purposes, i.e. to depict typical swimmer trajectories in the two-dimensional Cartesian space, the trajectory equations \eqref{eq:trajectory1} will be supplemented by an additional equation of motion along the $x$-axis, i.e., with 
\begin{equation}
\tilde{\dot{x}}(\tilde t) =  Pe_s \cos\theta(\tilde t) + Pe_f \tilde u_f\left(\tilde y(\tilde t)\right),
\label{eq:trajectory2}
\end{equation}
comprising the $x$-components of swimmer self-propulsion (first term) and advection with the flow. Figure \ref{fig:swing_tumble}a shows the trajectory  of a prolate swimmer with aspect ratio $\alpha=3$ (see Section \ref{subsec:Deterministic}), starting its motion with upstream orientation from high shear regions, e.g. near the lower channel wall; this swimmer performs tumbling motion. However, when another, identical, prolate swimmer starts its upstream swimming motion near the midchannel line, it performs swinging motion. 

Oblate swimmers ($\alpha=1/3$) with the same initial near-wall position and upstream swimming orientation, as seen in Fig. \ref{fig:swing_tumble}b, could also perform tumbling motion. When compared with prolate swimmers ($\alpha=3$) with the same swimming P\'eclet number, oblate swimmers have a greater lateral displacement during the tumbling motion; this becomes clear by comparing the swimming trajectories of the prolate and oblate spheroidal swimmers of panels a and b of Fig. \ref{fig:swing_tumble}, respectively.  The oblate swimmer starting its upstream swimming motion closer to the midchannel line, as shown in Fig. \ref{fig:swing_tumble}b performs swinging motion in the middle of the channel, with more lateral migration relative to the prolate swimmers, similar to the respective tumbling motions. 

The aforementioned difference of lateral migration is caused by the tendency of prolate swimmers to move in upstream and downstream directions in shear flow, i.e. $\theta=\pi$ and $\theta=0$, respectively; while sheared oblate swimmers are more likely to swim upwards and downwards, i.e. $\theta=\pi/2$ and $\theta=-\pi/2$, respectively (see Ref. \cite{MRSh1}). The prevalence of sheared prolate swimmers to swim parallel to the flow streamlines leads to {\em shear-trapping}; this effect entails the prolate swimmers to swim  predominantly parallel to the flow in high shear, near wall regions, and preventing them from reaching the middle parts of the channel, as discussed in Section \ref{susubsec:NonChiralProb}. Thus, suspensions of prolate swimmers, when subject to Poiseuille flow, exhibit {\em midchannel depletion} \cite{Rusconi2014,ZottlPRL2012,ZottlPoiseuille2013,Ezhilan}. However, as Section \ref{susubsec:NonChiralProb} laid out, oblate swimmers evacuate the high shear regions near the walls, once their long axes are aligned with the flow, and swim toward the low shear regions in the middle of the channel. Passing the midchannel line toward the other side, oblate swimmers are again subject to shear-induced rotation; we term this behavior of oblate swimmers, when subject to Poiseuille flow, as {\em shear-escaping}, which has not been reported in previous studies.

\begin{figure*}[t!]
	\begin{center}
		\begin{minipage}[b]{0.3\textwidth}
			\centering
			\includegraphics[height=0.7\linewidth]{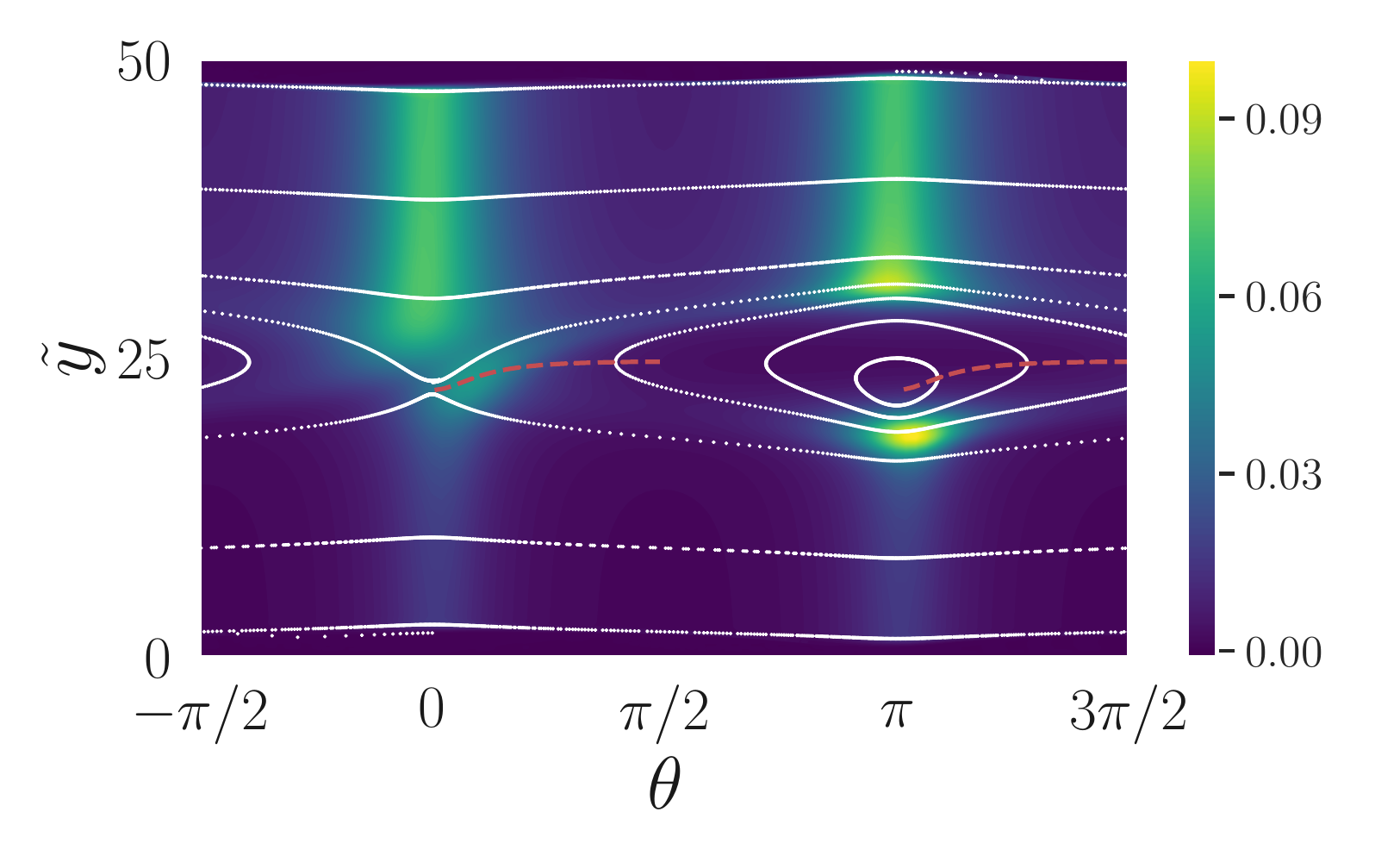}
			\vskip-2mm{(a) $Pe_f=200$ and $\Gamma=5$.}\label{chiralpoiseuillep1h}
		\end{minipage}\hskip4mm
		\begin{minipage}[b]{0.3\textwidth}
			\begin{center}
				\includegraphics[height=0.7\linewidth]{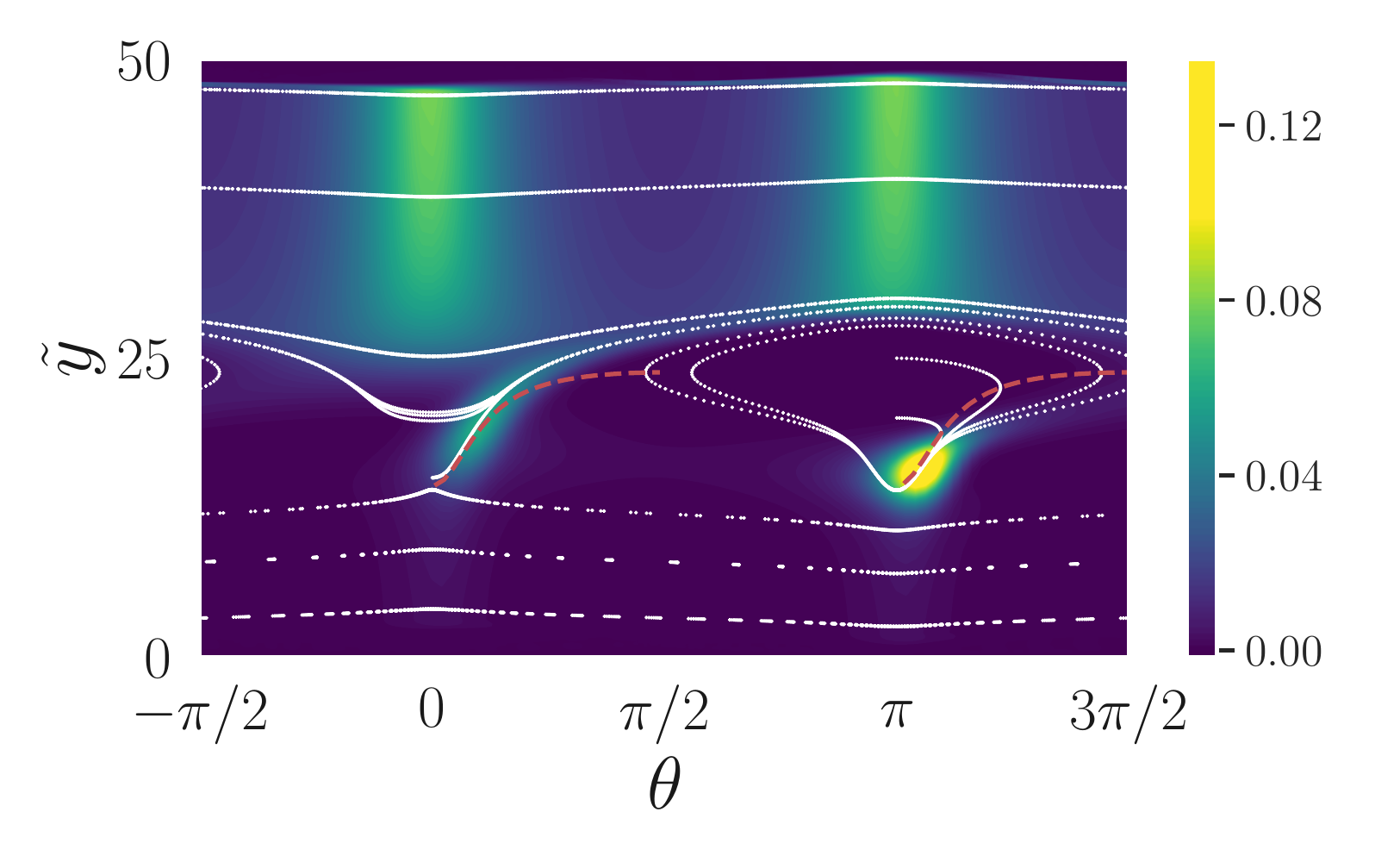}
				\vskip-2mm{(b) $Pe_f=200$ and $\Gamma=20$.}\label{chiralpoiseuillep4h}
			\end{center}
		\end{minipage}\hskip4mm
		\begin{minipage}[b]{0.3\textwidth}
			\centering
			\includegraphics[height=0.7\linewidth]{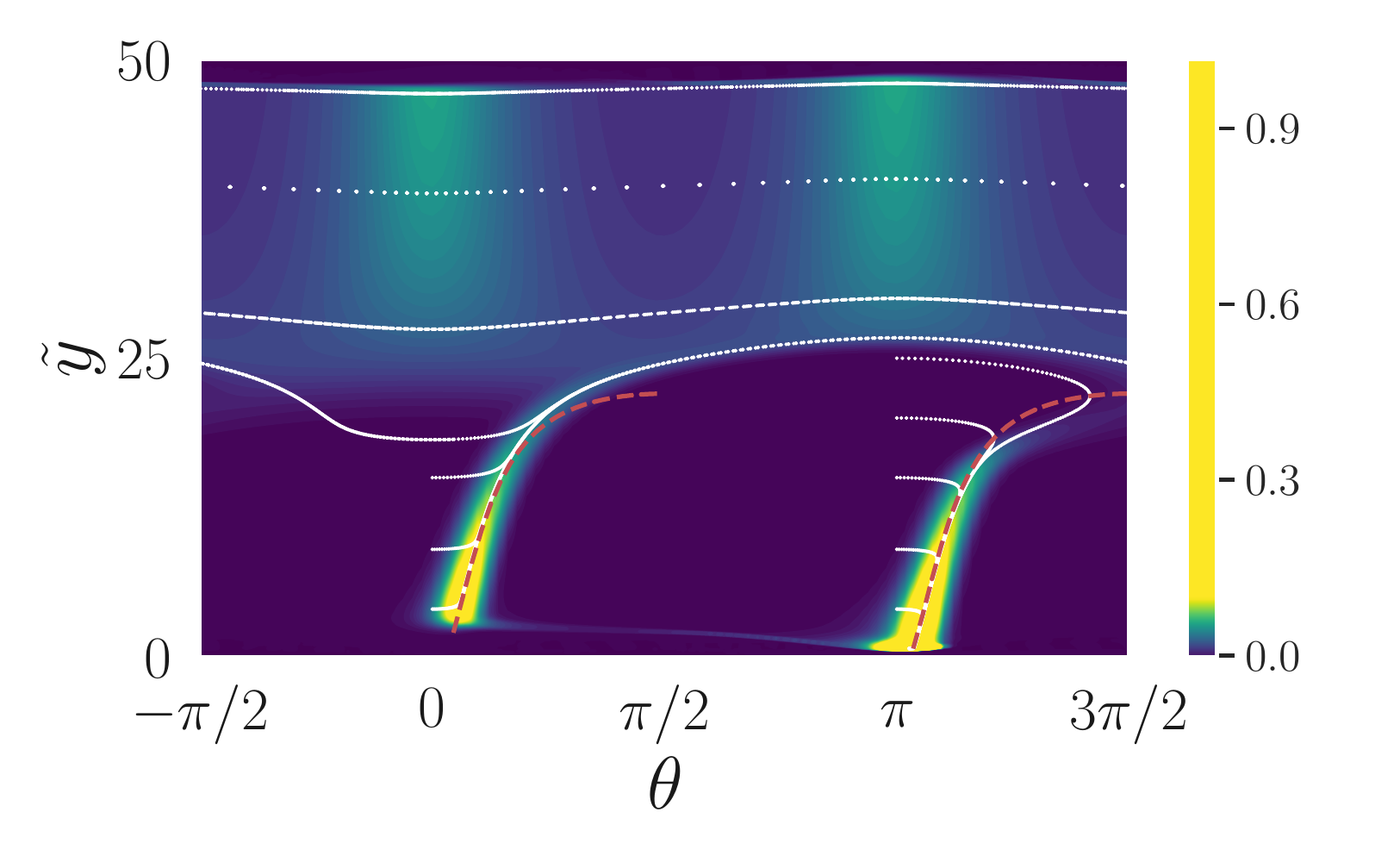}
			\vskip-2mm{(c) $Pe_f=200$ and $\Gamma=50$.}\label{chiralpoiseuillep2h}
		\end{minipage}\\
		\begin{minipage}[b]{0.3\textwidth}
			\begin{center}
				\includegraphics[height=0.7\linewidth]{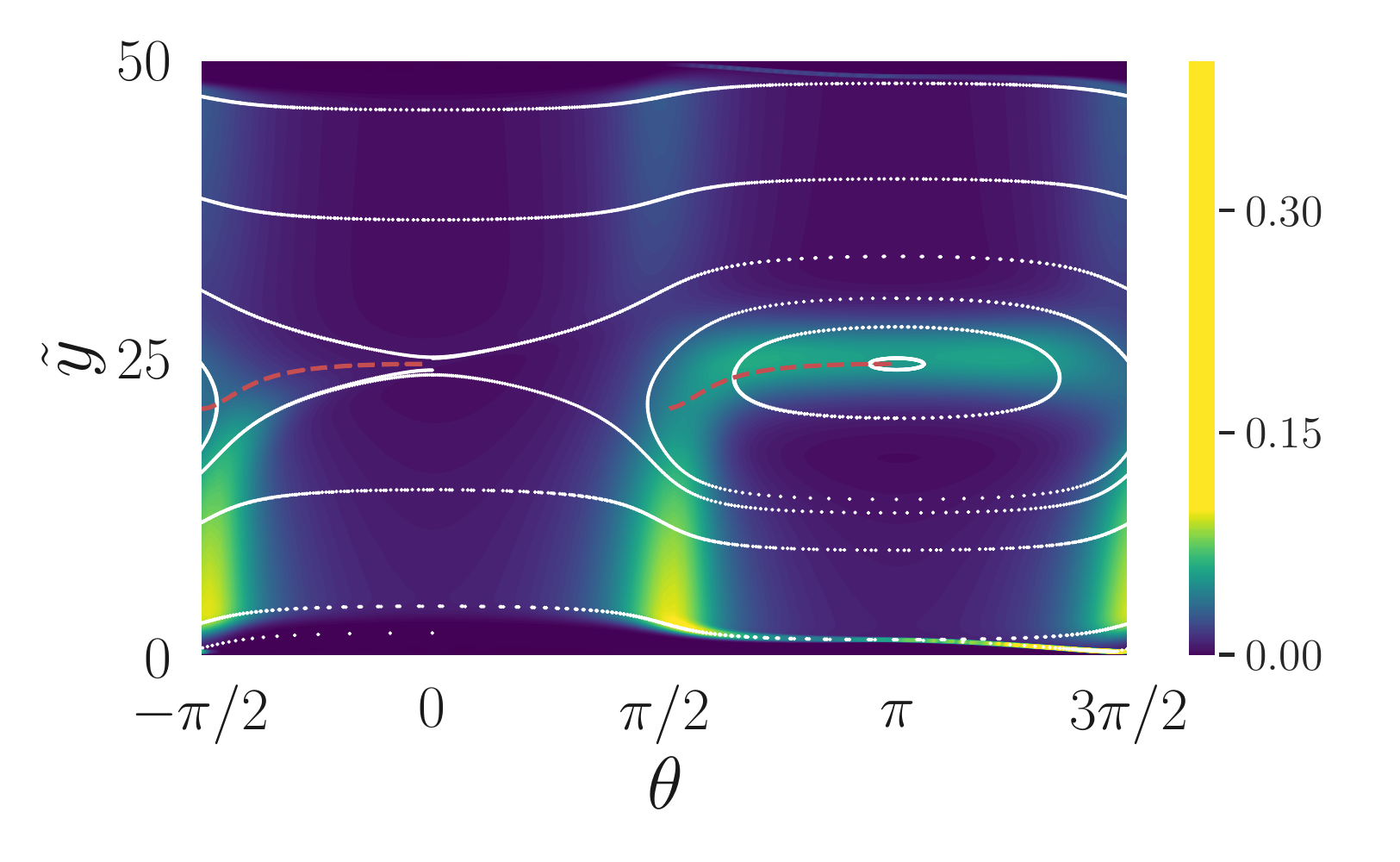}
				\vskip-2mm{(d) $Pe_f=200$ and $\Gamma=5$.}\label{chiralpoiseuilleo5}
			\end{center}
		\end{minipage}\hskip4mm
		\begin{minipage}[b]{0.3\textwidth}  
			\begin{center}
				\includegraphics[height=0.7\linewidth]{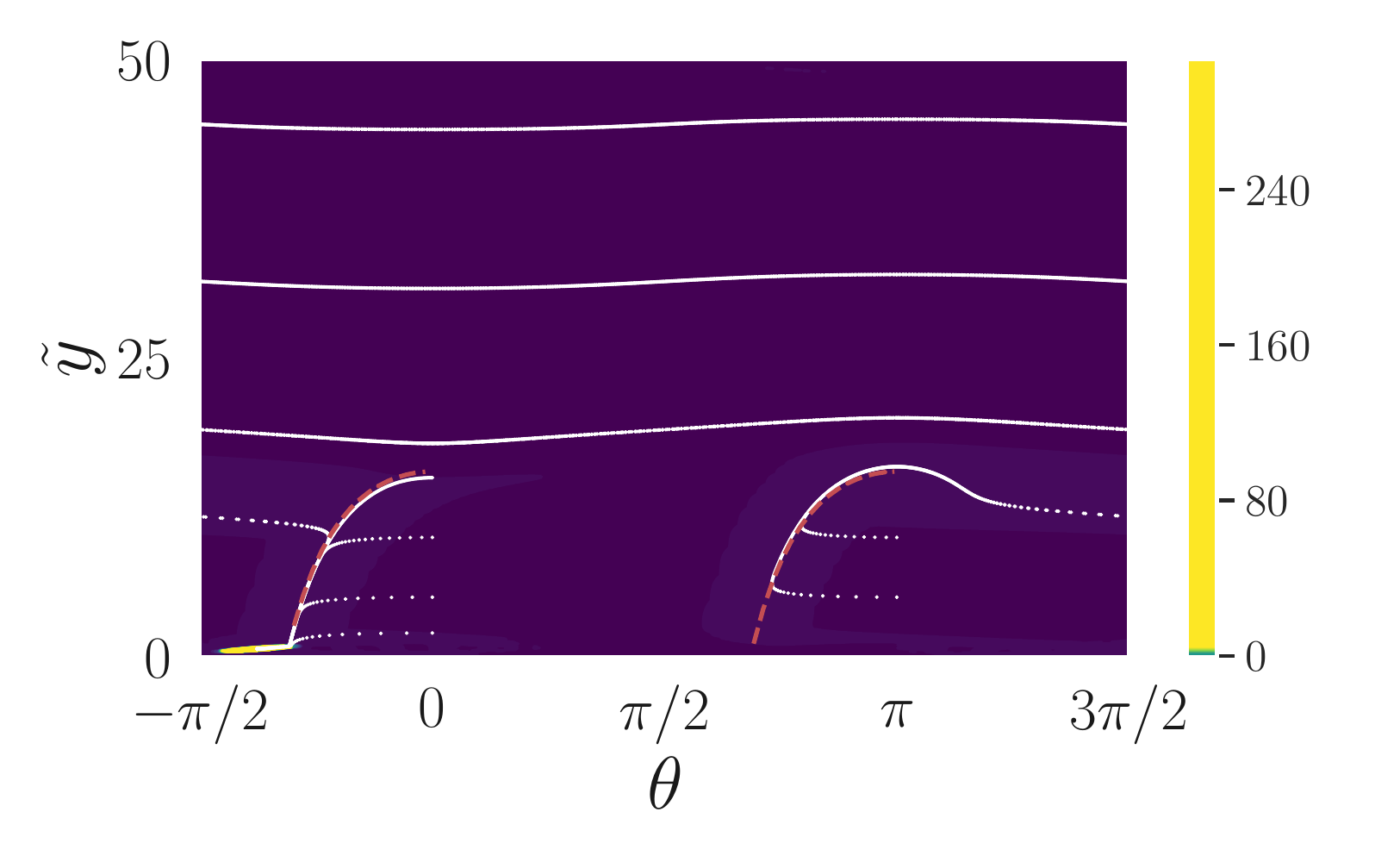}
				\vskip-2mm{(e) $Pe_f=200$ and $\Gamma=25$.}\label{chiralpoiseuilleo7}
			\end{center} 
		\end{minipage}\hskip4mm
		\begin{minipage}[b]{0.3\textwidth}
			\begin{center}
				\includegraphics[height=0.7\linewidth]{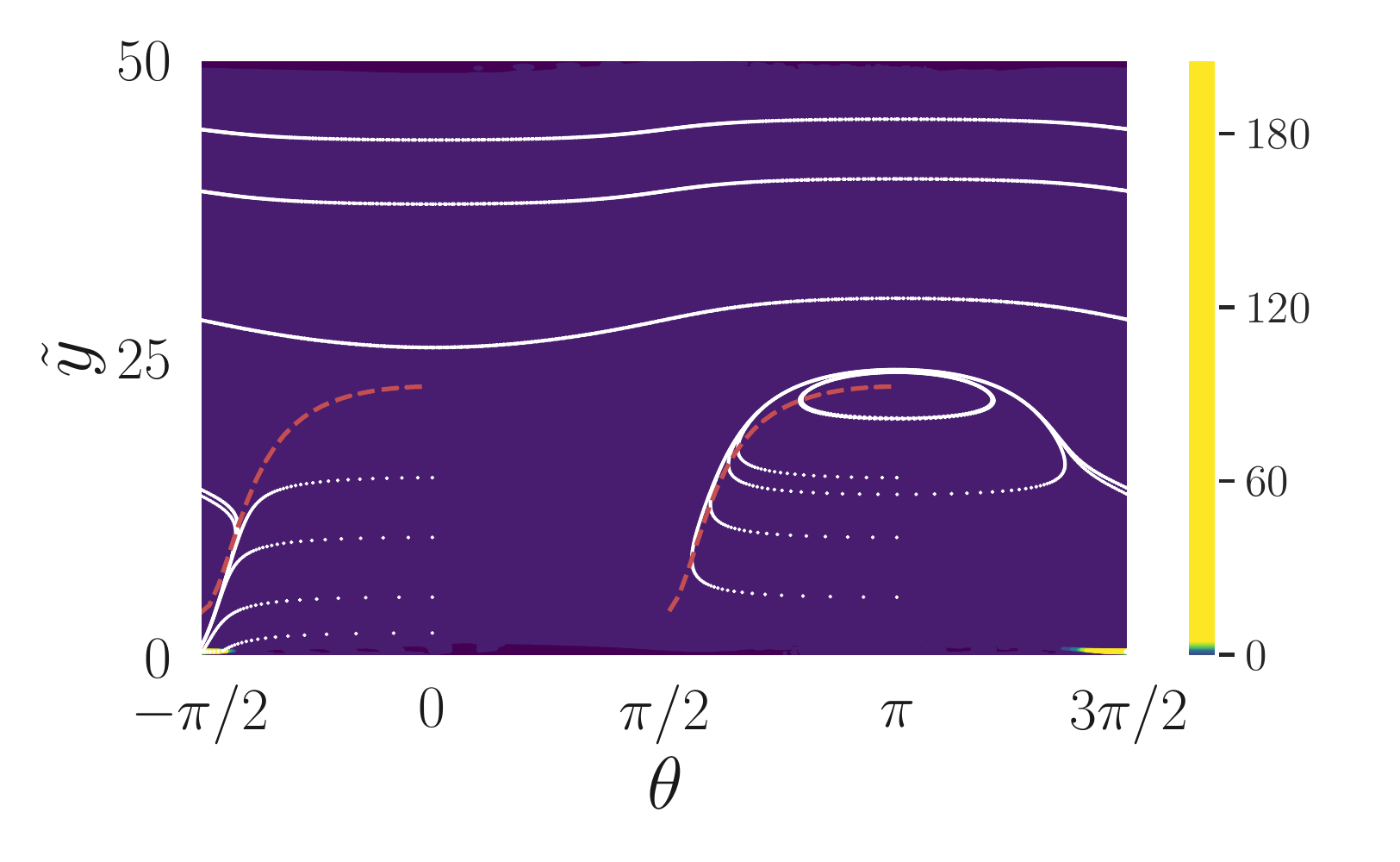}
				\vskip-2mm{(f) $Pe_f=200$ and $\Gamma=100$.}\label{chiralpoiseuilleo8}
			\end{center}  
		\end{minipage}
	\end{center}
	\vskip-2mm
	\caption{Shear-dominant rotational behavior and migration of chiral spheroidal swimmers with the swimming P\'eclet number $Pe_s = 50$, with different chirality coefficients, $\Gamma$, in Poiseuille flow with $Pe_f = 200$. Panels a--c show the PDF, $\tilde\Psi(\tilde y , \theta)$, of prolate swimmers of aspect ratio $\alpha=3$, and d--f show $\tilde\Psi(\tilde y , \theta)$ for chiral oblate swimmers of aspect ratio $\alpha=1/3$. The deterministic swimmer trajectories are shown with white dotted lines and the red dashed lines represent the pinning curves.
	} 
	\label{fig:ShearDominatedPoiseuille}
\end{figure*}

\subsection{Chiral swimmers in Poiseuille flow}
\label{SubSec:ChiralPoiseuille}

In absence of shear flow, chiral swimmers traverse circular trajectories \cite{Teeffelen:PRE2008,Palacci2013,Bechinger:PRL2013,QuinckeRotor2019, Teeffelen, Reichhardt:2013, Xue:EPL2015, Mijalkov:2015, Wykes_2016, Lowen:EPJST2016,  Liebchen_2016}. The dimensionless parametrization of swimming velocity, geometry and chirality of a sample chiral swimmer is carried out in Appendix \ref{app:parameters}. When chiral swimmers are subject to shear flow,  their rotational dynamics is determined by the interplay between the shear- and chirality-induced angular velocities. The chirality-induced angular velocity is constant and independent of  the swimmer orientation, $\theta$, while the shear-induced angular velocity is a function of $\theta$ (see Eqs. \eqref{eq:w_ext} and  \eqref{eq:w_f}). Combination of the two aforementioned contributions to angular velocity of chiral swimmers gives rise to the probabilistic behavior that is discussed below.

Figures \ref{fig:ShearDominatedPoiseuille}a-f show the PDF of prolate (a-c) and oblate (d-f) chiral swimmers with aspect ratios $\alpha=3$ and $\alpha=1/3$, respectively, subject to Poiseuille flow with P\'ecelt number $Pe_f = 200$. Figures \ref{fig:ShearDominatedPoiseuille}a and b show that prolate chiral swimmers with $\Gamma=5$ and $\Gamma=20$ are split into two downstream and upstream swimming subpopulations in the upper half and their PDF does not change significantly as compared  to the case of nonchiral prolate swimmers in Fig. \ref{fig:pef_poiseuille}c. In the lower half of the channel, however, the same prolate chiral swimmers accumulate around the point of phase space with orientation $\theta = \pi$, as well as, along the the red dashed lines that are addressed below. 

This probabilistic pattern can be explained through deterministic stability analysis of swimmer trajectories. Indeed, as laid out in Appendix \ref{subsec:MidChannelFixedPoints}, prolate chiral swimmers with $\Gamma<\Gamma_\ast(0)$ (corresponding to chiral swimmers in Figs. \ref{fig:ShearDominatedPoiseuille}a and b with $\Gamma=5$ and $\Gamma=20$, respectively) have two {\em off-centered fixed points}: a center at ($y_\ast , \pi$) and a saddle-point at ($y_\ast, 0$) in the phase space; the PDF accumulates around the center while it is removed from the areas around the saddle point and inside the separatix. The red dashed lines in all panels of Fig. \ref{fig:ShearDominatedPoiseuille}, including panels a and b, represent two of the {\em nullclines}  \cite{Nayfeh2008} formed by the vanishing of total angular velocity, i.e., $\tilde\omega =0$ (see Eq. \eqref{eq:FPeq2}). These particular nullclines, termed {\em pinning curves} \cite{MatsunagaPRL2017,MRSh2}, represent the sequence of stable swimmer orientations across the latitudes $\tilde y$, i.e., values of $\theta$ that further satisfy   $\partial \tilde\omega/\partial \theta <0$ and are given as $\theta_p(\tilde y)$ (see Appendix \ref{subsec:Pinning}).

Figure \ref{fig:ShearDominatedPoiseuille}c shows that prolate chiral swimmers with $\Gamma=50$ have less pronounced  downstream and upstream subpopulations of swimmers in the upper half of the channel, due to the added contribution of chirality-induced  and shear-induced angular velocities. The PDF of chiral swimmers concentrate along the pinning curves that now reach the lower wall. Particularly, accumulation of PDF around the point where the pinning curve touches the lower wall with the swimming orientation $\theta \simeq \pi$ is very significant. The emergence of this {\em near-wall stable fixed point} is limited to the range of swimmer chirality coefficients $\Gamma_\ast(0)\leq\Gamma\leq\Gamma_{\ast\ast}(0)$, as described in Appendix \ref{subsec:NearWallFixedPoints}. In analytical terms, the PDF of this class of swimmers is particularly enhanced around the stable near-wall fixed point ($\tilde y_\ast, \theta_\ast$) at $\pi\leq\theta_\ast <3\pi/2$, where the pinning curve meets the lower wall, as explained in Appendix \ref{subsec:Pinning}. The lower bound for existence of stable near-wall fixed point coincides with the lower bound for orientational pinning of prolate swimmers near the lower wall, i.e., $\Gamma_w \simeq\Gamma_\ast(0)$; this is deduced by comparing Eqs. \eqref{eq:LWFPThresh} and \eqref{GammaAstPosieuille}, after a small correction due to the role of steric effects.

Figure \ref{fig:ShearDominatedPoiseuille}d shows that oblate chiral swimmers with $\Gamma\ll\Gamma_\ast(0)$ accumulate around the center, i.e.  $(\tilde y_\ast, \pi)$, and their PDF in the upper half of the channel is depleted. However, as Fig. \ref{fig:ShearDominatedPoiseuille}e shows, oblate swimmers with higher chirality coefficients still falling in the range $\Gamma<\Gamma_\ast(0)$ migrate toward the lower half of the channel; their PDF is completely accumulated near the lower wall ($\tilde y/\tilde H \simeq 0 $) with the swimming orientation $\theta \simeq \theta_\ast (\tilde y)$ that points into the wall. Figure \ref{fig:ShearDominatedPoiseuille}f shows that oblate chiral swimmers with $\Gamma_\ast(0)\leq\Gamma\leq\Gamma_{\ast\ast}(0)$ possess a stable near-wall fixed point ($\tilde y_\ast, \theta_\ast$) at $-\pi/2<\theta_\ast \leq0$ around which their PDF is concentrated. A quick comparison between  Eqs. \eqref{eq:LWFPThresh} and \eqref{GammaAstAstPosieuille} reveals that for oblate swimmers, the lower bound for existence of stable near-wall fixed point coincides with the upper bound for orientational pinning near the lower wall, i.e. $\Gamma_w\simeq\Gamma_{\ast\ast}(0)$.

Prolate and oblate swimmers with chirality coefficient $\Gamma>\Gamma_{\ast\ast}(0)$ exhibit continuous rotational motion, since they have nonvanishing angular velocity, i.e., $\tilde\omega\neq 0$ in both halves of the channel. Thus, we term this regime of behavior as {\em chirality dominant swimming}; since the PDF is nearly isotropic in this regime (not shown), we do not discuss it further.

\subsubsection{Summary: Migration patterns of chiral swimmers}
\label{subsec:MigrationPatterns}

The probabilistic results of Section \ref{SubSec:ChiralPoiseuille} show that prolate and oblate chiral swimmers with the same chirality coefficient, despite having similar orientational dynamics when adjusted for the $\pi/2$ shift in their orientation relative to the flow direction, exhibit different migration patterns in shear dominant regime. The results of deterministic analysis of Appendix  \ref{App:ChiralStability} and probabilisitic results of  Section \ref{SubSec:ChiralPoiseuille} can be summarized for prolate chiral swimmers ($\alpha>1$) as,
\begin{align}
\left\{\begin{array}{ll}
\triangleright\,\,&\textrm{Off-centered fixed point:}\\
&\quad\Gamma < \Gamma_{\ast}(0),\\ \\
&\bullet\quad\textrm{Center and saddle-point in the lower half,}\\
&\bullet\quad \textrm{Accumulation of PDF around center}\\ 
&\quad \quad\textrm{and pinning curve,}\\
&\bullet\quad \textrm{Depletion of other parts of the lower half}\\
\\
\triangleright\,\,&\textrm{Stable near-wall fixed point:}\\
&\quad\Gamma_{\ast}( 0)\leq \Gamma \leq \Gamma_{\ast\ast}( 0) ,\\
&\quad\Gamma_w \simeq  \Gamma_{\ast}( 0) ,\\ \\
&\bullet\quad\textrm{Accumulation of PDF around} \\
&\quad\quad\textrm{the center and pinning curve}\\
\\
\triangleright\,\,&\textrm{Chirality-dominant swimming:}\\
&\quad\Gamma > \Gamma_{\ast\ast}(0).
\end{array}\right.
\label{eq:summProlate}
\end{align}
While for oblate chiral swimmers ($\alpha<1$), it can be summarized as,
\begin{align}
\left\{\begin{array}{ll}
\triangleright\,\,&\textrm{Off-centered fixed point:}\\
&\quad\Gamma < \Gamma_{\ast}(0),\\ \\
&\bullet\quad \textrm{Accumulation around the center/}\\\
&\quad\quad \textrm{near-wall accumulation,}\\
&\bullet\quad \textrm{Depletion of the upper half}\\
\\
\triangleright\,\,&\textrm{Stable near-wall fixed point:}\\
&\quad\Gamma_{\ast}( 0)\leq \Gamma \leq \Gamma_{\ast\ast}( 0) ,\\
&\quad\Gamma_w \simeq  \Gamma_{\ast\ast}( 0) ,\\ \\
&\bullet\quad\textrm{Complete near-wall accumulation}\\
\\
\triangleright\,\,&\textrm{Chirality-dominant swimming:}\\
&\quad\Gamma > \Gamma_{\ast\ast}(0).
\end{array}\right.
\label{eq:summOblate}
\end{align}
In Section \ref{SubSubSec:Separation}, we discuss some of the implications and potential applications for the above categorization.  As we pointed out in Section \ref{subsec:Smoluchowski_eq}, near-wall behavior must be regarded as a preliminary result that warrants further studies of the near-wall motion of swimmers, where steric and hydrodynamic effects become dominant.

\begin{figure}[t!]
	\includegraphics[width=0.95\textwidth]{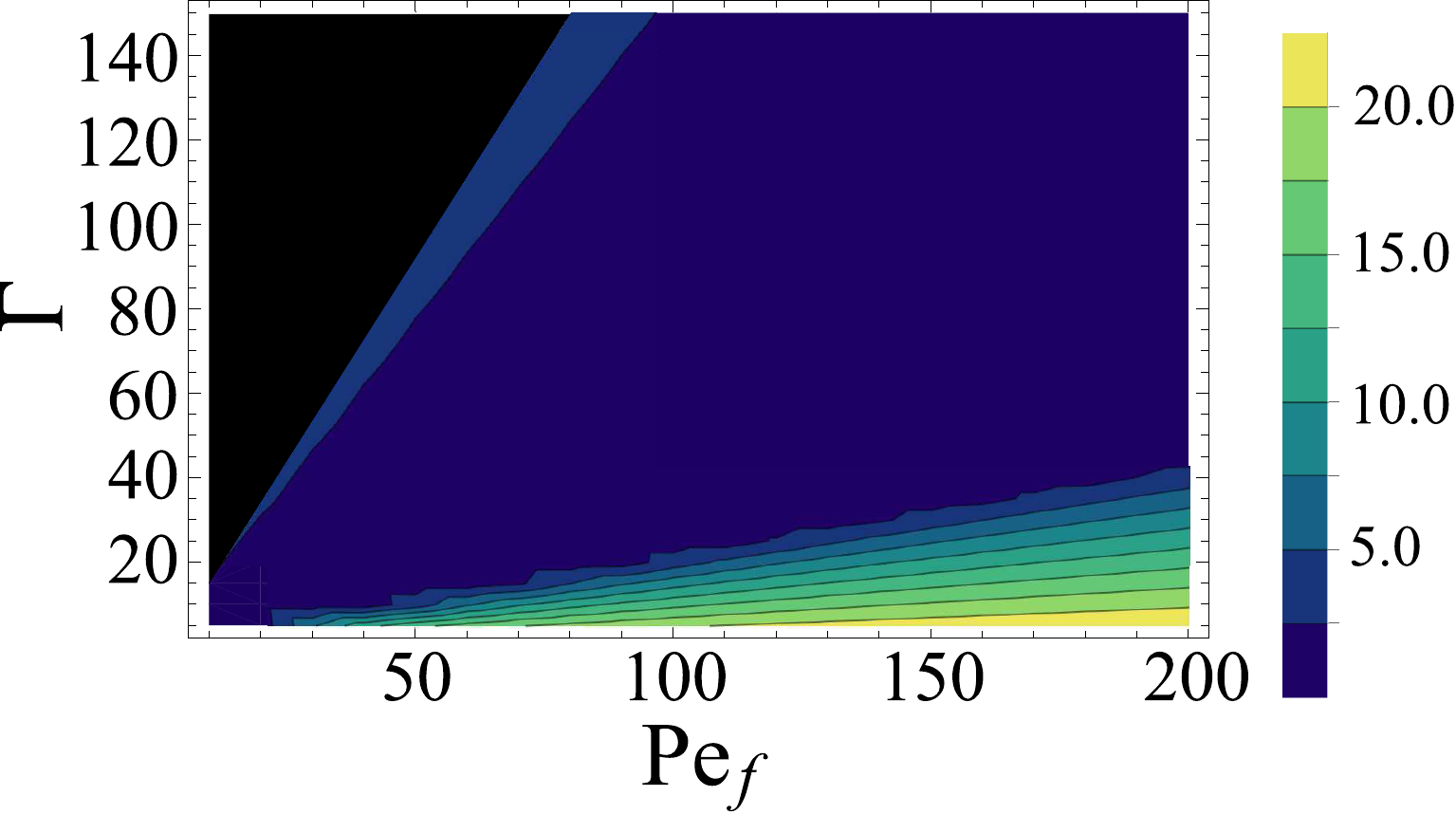}
	\vskip\baselineskip\caption{ Color map of off-centered fixed point (center) latitude, $\tilde{y}_\ast$, of prolate swimmers with aspect ratio $\alpha=3$ and swimming P\'eclet number $Pe_s=50$ with a range of chirality coefficients $\Gamma$, in Poiseuille flow with channel height $\tilde H = 50$ and a range of flow P\'eclet numbers $Pe_f$. The black triangular area in the top left of phase diagram shows the chirality-dominant regime with no orientational pinning. Focusing latitudes higher than $\tilde{y}_\ast \approx5$ represent midchannel focusing.}
	\label{fig:FocusingLatitudeProlates}
\end{figure}

\subsubsection{Difference in migration patterns: potential routes to separation of chiral swimmers}
\label{SubSubSec:Separation}

In this section we suggest a possible route for separation of chiral swimmers with different chirality coefficients, $\Gamma$. 
Figure \ref{fig:FocusingLatitudeProlates} indicates the off-centered fixed point (center) latitude, $\tilde{y}_\ast$ for prolate swimmers ($\alpha = 3$) with $Pe_s=50$; the range of colors represent the range of focusing latitudes across the lower channel half. Prolate swimmers with low ratios of $\Gamma/Pe_f$ (bottom-right corner of Fig. \ref{fig:FocusingLatitudeProlates}) accumulate around the center, located closer to channel center. As summarized in Section \ref{subsec:MigrationPatterns}, for these chiral swimmers, the lower half of the channel is depleted and the PDF mostly accumulates around this center. However, prolate swimmers with higher ratios of $\Gamma/Pe_f$, posses a stable near-wall fixed point and accumulate closer to the wall. Chiral swimmers with ratios $\Gamma/Pe_f\gg1$ (not shown) fall within chirality-dominant regime and do not experience pinning or focusing at any part of the channel. These differences can be utilized for chirality-based separation of chiral swimmers with similar motilities and shapes.

\section{Concluding Remarks}
\label{Sec:Conclusion}

We formulate the Smoluchowski equation governing the probability distribution function of both chiral and nonchiral spheroidal swimmers, subjected to Poiseuille flow inside a two dimensional microchannel with flat walls. We solve the Smoluchowski equation by finite element simulations. The resulting probability distribution function (PDF) in the latitude-orientation phase space is then analyzed with the help of the corresponding deterministic trajectories accompanied by the linear stability analysis. These results are then summarized as the formation of different classes of fixed points contribute to different migration patterns of nonchiral, as well as, chiral swimmers in pressure driven (Poiseuille) flow.

In the case of nonchiral swimmers, we establish that prolate swimmers get trapped in high shear regions (shear-trapping), while oblate swimmers evacuate these areas (shear-escaping). In other words, prolate swimmers evacuate the separatix while oblate swimmers occupy the $\tilde y-\theta$ phase space inside it.

Chiral swimmers in Poiseuille flow behave differently from nonchiral swimmers due to their intrinsic, approximately constant, angular velocity (circle swimming). When subjected to variable shear rate (associated with Poiseuille flow) across the channel, depending on their chirality coefficients, prolate and oblate swimmers fall within different regimes of rotational dynamics, corresponding to specific segments across the channel height. The stability analysis of swimmer trajectories within the $\tilde y-\theta$ phase space also reveals different categories of fixed points for chiral swimmers in Poiseuille flow with different $\Gamma/Pe_f$ ratios. 

Depending on the latitude across the channel, $\tilde y$, and chirality coefficient $\Gamma$, swimmers may exhibit either Jeffery orbits, modified by chirality of swimmers, or pinning (stable orientations). These dynamics are further complicated by different placements of fixed points for chiral swimmers depending on the ratio of $\Gamma/Pe_f$, emphasizing the role of stability analysis of the said fixed points. The above factors lead to different translational dynamics, such as accumulation of chiral swimmers around an off-centered fixed point (center) and the pinning curve, depletion of lower (upper) half of the channel for prolate (oblate) chiral swimmers and accumulation of oblate swimmers with higher chirality coefficients near the lower wall. Swimmers with very high chirality coefficients, fall within the chirality-dominant regime; thus, their persistence length of swimming shrinks and they exhibit a nearly uniform PDF in the $\tilde y - \theta$ domain within the channel. Based on the aforementioned differences in translational dynamics of chiral swimmers, we suggest a possible route for chirality-based separation of chiral swimmers with similar motility and aspect ratios. 


The results of this study, may be applied to the artificial, L-shaped, chiral swimmers of Ref. \cite{Bechinger:PRL2013} (see  Appendix \ref{app:parameters}), with certain precautions. First, the shear-induced rotations of these swimmers are assumed, within our model, to be approximated by the shear-induced rotations of spheroidal bodies; future studies could be designed to take the real geometries into account. Secondly, due the self-phoretic mechanism of propulsion, the chirality-induced angular velocity of these swimmers is proportional to their swimming velocity \cite{Bechinger:PRL2013}; as such, their radius of rotational orbits, in still fluids, are fixed. Therefore, depending on the height of channel and flow strength, these particular swimmers might fall within, either, swimming-dominant or shear-dominant regimes; consequently, in Poiseuille flow, their rotational dynamics  in the lower half of the channel could either range from pinning to chirality-dominant, or from modified Jeffery orbits to pinning regimes, respectively. Thirdly, since the L-shaped, chiral swimmers of Ref. \cite{Bechinger:PRL2013} have two different senses of rotation, the two species will behave differently, under the same flow conditions. For example, swimmers with $\Gamma > 0$ will experience pinning and have their fixed points in the lower half of the channel; however, the ones with $\Gamma < 0$ will experience these effects in the upper half. 

It is worth noting that inclusion of hydrodynamic interactions with channel walls (see Ref. \cite{ZottlPRL2012} concerning nonchiral swimmers and our earlier work for passive spheroids \cite{MRSh2}), and the details of steric interactions, could significantly change the results that are discussed in this paper. The results of this paper must be seen as a generalization of the results of Ref. \cite{ZottlPoiseuille2013} to oblate and chiral swimmers. As such, any generalization of the predictions of this paper to real systems must be limited to the results concerning the behavior of swimmers away from the walls.

\section{CONFLICT OF INTEREST}
The authors have no conflicts to disclose.

\section{AUTHOR CONTRIBUTIONS}
A.A. and A.P. performed the numerical simulations and M.R.S. conducted the deterministic analysis. All authors contributed to the discussions and prepared the initial draft of the manuscript. M.R.S. and A.N. finalized the discussions. M.R.S. wrote the final manuscript. A.A. and A.P.  contributed equally to this work. A.N. conceived the study and supervised the research. 

\section{DATA AVAILABILITY}
The data that support the findings of this study are available within the article.

\appendix

\section{Choice of parameter values}
\label{app:parameters}

In this paper, the dimensionless parameter values  are within the ranges $Pe_s=50-100$, $Pe_f=0-300$, $\Gamma=0-150$, $\tilde H=20-50$; as we will show below for a self-phoretic active particle, as an example, these values fall within experimentally realizable ranges of parameters for synthetic active particles. 
For  L-shaped active particles with self-phoretic active motion (see Refs. \cite{Bechinger:PRL2013,Lowen:EPJST2016}) with the swimming velocity $V_s \simeq 1.25\, \mu{\mathrm{m}}\cdot{\mathrm{s}}^{-1}$ and   the intrinsic angular velocity $\omega_c \simeq 0.2\,{\mathrm{s}}^{-1}$ taken from Ref.
\cite{Bechinger:PRL2013}. We use the geometrical specifications of Ref. \cite{Bechinger:PRL2013} to approximate the active particles as spheroidal particles; hence the aspect ratio of the L-shaped particle can be taken as $\alpha=3$ and radius of the reference sphere can be approximated as $R_{\mathrm{eff}} \simeq  5 \, \mu{\mathrm{m}}$, with Stokes rotational diffusivity of $D_{0R} =  1.3 \times10^{-3}\,{\mathrm{s}}^{-1}$. Therefore, we get the swimming P\'eclet number $Pe_s \simeq 192$ and the chirality coefficient $\Gamma = \omega_c / D_{0R}\simeq 154$. In this case, our choices of values for $Pe_f$ and $\tilde H$ are mapped to shear rates of $\dot\gamma\simeq 0 - 0.39\,{\mathrm{s}}^{-1}$ and actual channel heights of  $H\simeq 100- 250\, \mu$m, respectively. The Reynolds number for such an active particle in such a channel would be $\text{Re} = \rho \dot\gamma H R_{\text{eff}} / \eta \simeq 4.8\times 10^{-3} $ at its maximum, falling well within the low Reynolds number regime of hydrodynamics.

We must emphasize that our model addresses the generic aspects of motion of chiral (circle swimming) active particles and overlooks specifics of shape and self-propulsion mechanism, which need to be included for quantitative comparisons with any specific real system. In the specific case of  L-shaped active particles of Ref. \cite{Bechinger:PRL2013}, the modification of Jeffery orbits due to the non-spheroidal shape must be taken into account. The values cited above are meant only as a gauge of swimming parameters.

\section{Particle-wall steric interaction}
\label{app:y0}

In order to model the impenetrable channel walls, we use a phenomenological, harmonic potential to mimic the steric interaction between the spheroidal swimmers and the walls: 
\begin{eqnarray}
\phi(\theta,\tilde{y};\alpha)=&&\frac{\kappa}{2}\left(\tilde{y}_{0}(\theta;\alpha)-\tilde{y}\right)^{2}\;\Theta\left(\tilde{y}_{0}(\theta;\alpha)-\tilde{y}\right)\\
&&\hspace{-10mm}+\frac{\kappa}{2}\left(\tilde{y}-(\tilde{H}-\tilde{y}_{0}(\theta;\alpha)\right)^{2}\;\Theta\left(\tilde{y}-(\tilde{H}-\tilde{y}_{0}(\theta;\alpha))\right)\nonumber
\,,
\label{eq:harmonic}
\end{eqnarray}
where $\Theta(\tilde{y})$ is the Heaviside step function and $\tilde{y}_0$ is the lateral distance between the center of the spheroids and the lower wall.

The distance $\tilde{r}_0$ between the center of ellipsoid and the point of contact with the lower wall has two components  $\tilde{x}_0$ and $\tilde{y}_0$, parallel and perpendicular to the channel walls, respectively; these components for an ellipsoidal particle with the aspect ratio $\alpha$ and swimming orientation $\theta$ are given by
\begin{eqnarray}
\tilde{y}_0(\theta;\alpha)&=&\alpha^{2/3}\sqrt{\sin^{2}\theta+\alpha^{-2}\cos^{2}\theta}
\label{y0}
\,,\\
\tilde{x}_0(\theta;\alpha)&=&\frac{\partial}{\partial\theta} 	\;\tilde{y}_0(\theta;\alpha)
\,\cdot
\label{eq:x0y0}
\end{eqnarray} 
The perpendicular component of steric force due to excluded volume interaction, modeled by the steric harmonic potential of Eq. \eqref{eq:harmonic}, is given as
\begin{equation}
F_{\perp}^{(\mathrm{st})}(\tilde{y}) = -\frac{\partial}{\partial\tilde{y}} \phi(\theta,\tilde{y};\alpha)
\,\cdot
\label{eq:F_steric}
\end{equation}
Therefore, by using the anisotropic resistance functions, we obtain the induced steric velocity as
\begin{equation}
\tilde u_y^{\mathrm{(st)}}(\tilde{y})= - (\Delta_+(\alpha)-\Delta_-(\alpha)\cos 2\theta)\frac{\partial}{\partial\tilde{y}} \phi(\theta,\tilde{y};\alpha)
\,\cdot
\label{eq:u_steric}
\end{equation}
The induced steric angular velocity is calculated by using Eqs. \eqref{eq:F_steric} and  \eqref{eq:u_steric} as
\begin{equation}
\begin{aligned}
\tilde{\omega}^{(\mathrm{st})}(\tilde{y})=
&\,\Delta_{R}(\alpha)\tilde{\boldsymbol{r}}_0\times\boldsymbol{F}^{(\mathrm{st})}(\tilde{y}) \\
&\hspace{-14mm} =-\Delta_{R}(\alpha)\;\kappa\;\tilde{x}_0(\theta;\alpha)\bigg[\tilde{y}_0(\theta;\alpha)-\tilde{y}\bigg]\Theta(\tilde{y}_{0}-\tilde{y})\\
&\hspace{-10mm}-\Delta_{R}(\alpha)\;\kappa\;\tilde{x}_0(\theta;\alpha)\times \\
&\hspace{-10mm}\quad\bigg[(\tilde{H}-\tilde{y}_0(\theta;\alpha))-\tilde{y}\bigg]\Theta((\tilde{H}-\tilde{y}_0(\theta;\alpha))-\tilde{y})
\, \cdot
\label{Eq:Vsteric2}
\end{aligned}
\end{equation}

\section{Constants of motion: Relative dominance of shear flow and swimming in rotational dynamics}
\label{App:Cs}

In the special case of nonchiral swimmers, the deterministic {\em bulk} trajectories (i.e., those not coming into contact with the channel walls) in Poiseuille flow can be obtained, by using the deterministic flux velocities, as
\begin{eqnarray}
\frac{\tilde{\dot{y}}}{\dot{\theta}} = \frac{2 Pe_s \sin\theta}{Pe_f\left(1-2\tilde y/\tilde H\right)\left(\beta(\alpha) \cos2\theta-1\right)}. 
\label{eq:dynamicEqs}
\end{eqnarray}
Integration of the above equation gives the constants of motion \cite{ZottlPRL2012,ZottlPoiseuille2013} as
\begin{eqnarray}
	C=\frac{Pe_f}{2Pe_s}\bigg(\!\tilde{y}(\tilde t)\!-\!\frac{\tilde{y}^{2}(\tilde t)}{\tilde H}\bigg) 
		- f\left(\theta (\tilde t) \right) \,,
\label{ccpoiseuille}
\end{eqnarray}
where for prolate ($\beta>0$) and oblate ($\beta<0$) particles, respectively, we have
\begin{align}
f(  \theta) =
\left\{\begin{array}{ll}
		&\frac{\tanh ^{-1}\!\left(\!
		\sqrt{2\beta(\alpha)/ (\beta(\alpha)+1)}\cos\theta
		\right)}{\sqrt{2\beta(\alpha) (\beta(\alpha)+1)}} \,; \quad \alpha >1 \,, \\
		\\
		&\frac{\tan ^{-1}\!\left(\!
		\sqrt{-2\beta(\alpha)/ (\beta(\alpha)+1)}\cos\theta
		\right)}{\sqrt{-2\beta(\alpha) (\beta(\alpha)+1)}} \, ; \quad \alpha <1\,\cdot  \\
\end{array}\right.
\label{eq:f(costheta)}
\end{align}
Equation \eqref{ccpoiseuille} gives the constant of motion $C_0$ for which the separatix (see Section \ref{susubsec:NonChiralDet}) touches the channel walls. Indeed, $C_0$ can be calculated by putting ($\tilde H/2 , 0$) and ($\tilde 0 , \pi$) in place of ($\tilde y , \theta$) in Eq. \eqref{ccpoiseuille}; the two points designate the position of the saddle-point (see Section \ref{susubsec:NonChiralDet}) and the point of contact between the separatix and the lower wall, respectively. Equating the two resulting constants of motion, we get,
\begin{equation}
		Pe_f^\ast = 16\frac{ Pe_s}{ \tilde H} f(0) \,\cdot
\label{Pef0}
\end{equation}
Equation \eqref{Pef0} gives the critical flow strength $Pe_f^\ast$ (for fixed swimming velocity, swimmer aspect ratio and channel height), where for $Pe_f > Pe_f^\ast$ the separatix is pushed away from the walls, as we see in Section \ref{SubSec:Non}.

\section{Stability analysis of chiral swimmers in Poiseuille flow}
\label{App:ChiralStability}

We investigate the fixed points associated with chiral swimmers with $\Gamma > 0$, i.e., points within the ($\tilde y , \theta$) phase space that simultaneously satisfy the conditions \eqref{eq:FPeq1} and \eqref{eq:FPeq2}, concerning the deterministic translational and angular velocities. Equation \eqref{eq:FPeq2} only has solutions in the lower half of the channel, $0 \leq \tilde y < H/2$. In the forthcoming analysis, we investigate the conditions for existence of solutions for both Eqs. \eqref{eq:FPeq1} and \eqref{eq:FPeq2}, i.e., the conditions for existence of fixed points in the lower half of the channel and perform the corresponding stability analysis. 

\subsection{Stability of off-centered fixed points }
\label{subsec:MidChannelFixedPoints}

In the middle of the channel, the condition for existence of fixed point, i.e. \eqref{eq:FPeq1}, can only be satisfied when $\sin \theta =0$. The coordinates of the { \em off-centered fixed points} are thus determined as $(\tilde y_\ast,0)$ and $(\tilde y_\ast,\pi)$, where $\tilde y_\ast$ is calculated by replacing $\theta_\ast = 0,\,\pi$ in Eq. \eqref{eq:FPeq2}, as 
\begin{equation}
\frac{\tilde y_\ast}{\tilde H} = \frac{1}{2} - \frac{\Gamma }{Pe_f} \frac{\Delta_R}{1-\beta(\alpha)}
\cdot
\label{eq:MidChFPs}
\end{equation}
Equation \eqref{eq:MidChFPs} produces a threshold value for the chirality coefficient of swimmers where for $\Gamma < \Gamma_w$ the closest distance between center of swimmers and the lower wall as allowed by the steric potential does not coincide with $\tilde y_\ast$, i.e., $\tilde y_\ast > \tilde y_0$, where $\tilde y_0 \equiv \tilde y_0 (0,\alpha) =  \tilde y_0 (\pi,\alpha)$ (see Appendix \eqref{app:y0}); by using Eq. \eqref{y0} we have,
\begin{align}
\tilde y_0 =
\left\{\begin{array}{ll}
\alpha^{-1/3}\,; \quad\alpha>1 \,,&\\ \\
\alpha^{2/3} \,;\quad \alpha<1\,\cdot
\end{array}\right.
\label{eq:y0}
\end{align}
Therefore, the upper threshold of chirality coefficient for the existence of off-centered fixed points is given by
\begin{equation}
\Gamma_w = \left( \frac{1}{2} - \frac{\tilde y_0}{\tilde H}\right) \frac{(1-\beta) Pe_f}{\Delta_R}
\,\cdot
\label{eq:LWFPThresh}
\end{equation}
In Eq. \eqref{eq:LWFPThresh}, one can assume ${\tilde y_0}/{\tilde H} \ll 1$ and thus this term is dropped in future comparisons of $\Gamma_w$ with other parameters.

As stated in Section \ref{subsec:Deterministic}, the linear stability of any given fixed point with regard to small perturbations is determined by the eigenvalues of the Jacobian matrix,  $\lambda$, derived by the solution of Eq. \eqref{eq:lambdas}. The values of $\lambda$ corresponding to the aforementioned off-centered fixed points, $(\tilde y_\ast,0)$ and $(\tilde y_\ast,\pi)$, are equal to the eigenvalues at the central fixed points, $(\tilde H/2, 0)$ and $(\tilde H/2, \pi)$, respectively, given by Eq. \eqref{eq:MidChLamPi_nonchiral}; indeed, the same procedure of Section \ref{subsec:Deterministic} is followed to calculate both sets of eignevalues. Therefore, the off-centered fixed point at $(\tilde y_\ast,0)$ is a {\em saddle point }, while $(\tilde y_\ast, \pi)$ is a {\em center}. Similarly, the homoclinic orbit connecting the saddle-point $(\tilde y_\ast,0)$ back to itself is the separatix that separates the periodic rotations (swinging) around the aforementioned center, i.e. $(\tilde y_\ast, \pi)$, from the {\em modified Jeffery orbits} (tumbling) that span the whole $[-\pi/2,3\pi/2)$ orientation space (see Section \ref{SubSec:ChiralPoiseuille}).

\subsection{Stable near-wall fixed points }
\label{subsec:NearWallFixedPoints}

Swimmers with chirality coefficients $\Gamma = \Gamma_w$ have two fixed points in contact with the lower wall, i.e., $(\alpha^{-1/3} ,0)$ and $( \alpha^{-1/3},\pi)$, as shown in Section \ref{subsec:MidChannelFixedPoints}. Thus, the threshold $\Gamma_w$ corresponds to persistent longitudinal swimming near the lower wall (either in upstream or downstream orientations), in the absence of hydrodynamic interactions with the walls.

Prolate and oblate swimmers with higher chirality coefficients, i.e.,  $\Gamma > \Gamma_w$, never satisfy Eq. \eqref{eq:FPeq2} when $\sin\theta = 0$. Therefore, the condition \eqref{eq:FPeq1} for existence of fixed points can only be satisfied when swimmers swim into the wall as a result of steric interactions. The coordinates of resulting {\em near-wall fixed points} within the $\tilde y - \theta$ phase space are calculated by by (1) finding the swimmer orientations $\theta_\ast$ that satisfy Eq. \eqref{eq:FPeq2} regarding the rotational dynamics that  (2) must also satisfy the criterion  $\sin \theta_\ast < 0$ regarding the translational dynamics. Thus, the lateral coordinate of the near-wall fixed point is determined by the distance of penetration of swimmers that swim into the lower wall at the said orientation: 
\begin{equation}
 \tilde y_\ast = \tilde y_0(\theta_\ast,\alpha) +\frac{Pe_s}{\kappa}\sin \theta_\ast,
\label{eq:swim-steric_bal}
\end{equation}
where $ \kappa$ is the rescaled stiffness of the harmonic excluded volume force and $\tilde y_0 (\theta_p,\alpha)$ is the distance from the lower wall below which it comes into effect (see Appendix \ref{app:y0}). Since the swimmers swim into the wall, i.e., $\sin\theta_\ast < 0$, Eq. \eqref{eq:swim-steric_bal} gives $\tilde y_\ast < \tilde y_0 (\theta_\ast,\alpha)$.

The eigenvalues of the Jacobian matrix (see Eq. \eqref{eq:lambdas}) at each near-wall fixed point $(\tilde y_\ast,\theta_\ast)$ are then given as (see Section \ref{subsec:Deterministic}), 
\begin{eqnarray}
&&\lambda =  - \frac{1}{2}\left(\beta Pe_f \sin2\theta_a + \kappa \right) \nonumber \\ 
&&\quad\quad\pm \bigg[ \frac{1}{4} \left(\beta Pe_f \sin2\theta_a + \kappa \right)^2  \nonumber\\ 
&&\quad\quad- \kappa \beta Pe_f \sin2\theta_a +  2\Gamma \Delta_R \frac{Pe_s}{\tilde H} \cos\theta_a \bigg]^{\frac{1}{2}}
\cdot
\label{eq:NWLam}
\end{eqnarray}
For the near-wall fixed point $(\tilde y_\ast,\theta_\ast)$ to be {\em stable}, both eigenvalues given by Eq. \eqref{eq:NWLam} must have negative real parts, i.e., $\mathrm{Re}[\lambda] < 0$; this happens if $\kappa \gg Pe_s$ and    
\begin{equation}
\beta \sin2\theta_\ast > 0
\, ,
\label{eq:NWLamSC}
\end{equation}
Therefore, the stability of the aforementioned fixed point is independent of the details of steric interaction potential, as long as the rescaled stiffness of harmonic potential $\kappa$ is sufficiently large. In Section \ref{subsec:Pinning} we show that the stability criteria of near-wall fixed points, i.e., Eq. \eqref{eq:NWLamSC}, is the same condition as the stability of swimmer orientation.

\subsection{Pinning: Stabilized swimmer orientation}
\label{subsec:Pinning}

In this section, we investigate a pair of {\em nullclines} \cite{Nayfeh2008} within the  $\tilde y -\theta$ phase space that correspond to stable orientations of swimmers at any given latitude, regardless of the lateral swimming velocity. Thus, we term these nullclines as the {\em pinning curves} \cite{MatsunagaPRL2017,MRSh2} and $\theta_p(\tilde y)$ gives the sequence of pinning orientations. The pinning conditions, i.e. the condition for stabilized orientation of swimmers, are written as $\tilde \omega(\tilde y , \theta_p) =0$ and $\partial\tilde \omega/\partial\theta \big|_{(\tilde y , \theta_p)} < 0$. Using Eqs. \eqref{eq:Gamma_f} and \eqref{eq:Gamma_ext}, we know that under variable shear rate corresponding to Poiseuille flow (see Eq. \eqref{shear1}), the first pinning condition, i.e., $\tilde \omega(\tilde y , \theta_p) =0$, is only satisfied in the lower half of the channel ($\tilde y < \tilde H /2$) and only for the range of swimmer chirality coefficients $\Gamma_{\ast}(\tilde y)\leq\Gamma\leq\Gamma_{\ast\ast} (\tilde y)$, where we have defined the lower and upper bounds of chirality coefficient of swimmers for orientational pinning as,
\begin{align}
&\Gamma_{\ast}(\tilde y) = \left(1-2\frac{\tilde y}{\tilde H}\right) \frac{1-\lvert\beta(\alpha)\rvert}{2\Delta_R}Pe_f,
\,
\label{GammaAstPosieuille}
\\
&\Gamma_{\ast\ast} (\tilde y)= \left(1-2\frac{\tilde y}{\tilde H}\right) \frac{1+\lvert\beta(\alpha)\rvert}{2\Delta_R}Pe_f
\,,
\label{GammaAstAstPosieuille}
\end{align}
respectively.

The second pinning condition, i.e., $\partial\tilde \omega/\partial\theta \big|_{(\tilde y , \theta_p)} < 0$ takes the form
\begin{equation}
 \beta Pe_f \left(1-2 \frac{\tilde y}{\tilde H}\right) \sin 2 \theta_p > 0
\cdot
\label{PinningCondition}
\end{equation}
In the lower half of the channel ($\tilde y < \tilde H /2$), the pinning condition \eqref{PinningCondition} implies
\begin{equation}
\beta \sin 2 \theta_p > 0
\cdot
\label{PinningCondition2}
\end{equation}
This is the same as the stability condition for near-wall fixed points, i.e., Eq. \eqref{eq:NWLamSC}. Noting that $\beta>0$ for prolate swimmers and $\beta<0$ for oblate swimmers, Eq. \eqref{PinningCondition2} further implies
\begin{align}
\left\{\begin{array}{ll}
\sin 2 \theta_p >1\,; \quad\alpha>1 \,,&\\ \\
\sin 2 \theta_p <1 \,;\quad \alpha<1\,,
\end{array}\right.
\label{PinningCondition3}
\end{align}
Thus, Prolate and oblate swimmers  have two pinning orientations, satisfying the pinning conditions \eqref{PinningCondition2}, that lay in the range 
\begin{align}
\left\{\begin{array}{ll}
 0<\theta^{(1)}_p<\pi/2 \,, \,\,\,  \pi<\theta^{(2)}_p<3\pi/2 \,; \quad \alpha>1
 \,,&\\ \\
 -\pi/2<\theta^{(1)}_p<0\,, \,\,\,  \pi/2<\theta^{(2)}_p<\pi\,; \quad\alpha<1
\,,
\end{array}\right.
\label{eq:pinningOrientations}
\end{align}
respectively. 
The pinning angles $\theta^{(2)}_p$ and $\theta^{(1)}_p$, for prolate and oblate chiral swimmers, respectively, satisfy the second stability criterion for near-wall fixed points, i.e., $\sin \theta_\ast<0$, as laid out in Section \ref{subsec:NearWallFixedPoints}. Therefore, when the pinning curves reach to the lower wall, there is {\em exactly one} stable near-wall fixed point that will lead to accumulation of chiral swimmers at the lower wall. This regime of behavior for chiral swimmers is delimited by putting $\tilde y = y_0$ inside lower and upper limits in Eqs. \eqref{GammaAstPosieuille} and \eqref{GammaAstAstPosieuille}, respectively. This deterministic result is further confirmed by our probabilistic analysis, as laid out in Section \ref{SubSec:ChiralPoiseuille}.

\bibliography{Chiral_Swimmers_arXiv}

\end{document}